\documentclass[prd,twocolumn,amsmath,amssymb,floatfix,superscriptaddress,nofootinbib,preprintnumbers]{revtex4-1}

\usepackage[utf8]{inputenc}
\usepackage{graphicx}
\usepackage{amssymb}
\usepackage{amsmath}
\usepackage{bm}
\usepackage{color}
\usepackage{enumitem}
\usepackage[linktocpage=true]{hyperref} 
\hypersetup{
    colorlinks=true,       
    linkcolor=red,          
    citecolor=blue,        
    filecolor=magenta,      
    urlcolor=blue           
}
\usepackage[all]{hypcap} 

\usepackage{myterms}
\definecolor{darkgreen}{cmyk}{0.85,0.1,1.00,0}

\newcommand{\nl}{\\ \indent}

\begin{document}

\title{\Large Neutrino Mass Priors for Cosmology from Random Matrices}

\author{\large Andrew J. Long}
\affiliation{Kavli Institute for Cosmological Physics, University of Chicago, Chicago, Illinois 60637, USA.}

\author{\large Marco Raveri}
\affiliation{Kavli Institute for Cosmological Physics, University of Chicago, Chicago, Illinois 60637, USA.}
\affiliation{Institute Lorentz, Leiden University, PO Box 9506, Leiden 2300 RA, The Netherlands.}

\author{\large Wayne Hu}
\affiliation{Kavli Institute for Cosmological Physics, University of Chicago, Chicago, Illinois 60637, USA.}
\affiliation{Department of Astronomy \& Astrophysics, Enrico Fermi Institute, University of Chicago, Chicago, IL 60637.}

\author{\large Scott Dodelson}
\affiliation{Department of Physics, Carnegie Mellon University, Pittsburgh, Pennsylvania 15312, USA.}

\date{\today}

\begin{abstract}
Cosmological measurements of structure are placing increasingly strong constraints on the sum of the neutrino masses, $\Sigma m_\nu$, through Bayesian inference. 
Because these constraints depend on the choice for the prior probability $\pi(\Sigma m_\nu)$, we argue that this prior should be motivated by fundamental physical principles rather than the ad hoc choices that are common in the literature.
The first step in this direction is to specify the prior directly at the level of the neutrino mass matrix $M_\nu$, since this is the parameter appearing in the Lagrangian of the particle physics theory.  
Thus by specifying a probability distribution over $M_\nu$, and by including the known squared mass splittings, we predict a theoretical probability distribution over $\Sigma m_\nu$ that we interpret as a Bayesian prior probability $\pi(\Sigma m_\nu)$.  
Assuming a basis-invariant probability distribution on $M_\nu$, also known as the anarchy hypothesis, we find that $\pi(\Sigma m_\nu)$ peaks close to the smallest $\Sigma m_\nu$ allowed by the measured mass splittings, roughly $0.06 \eV$ ($0.1 \eV$) for normal (inverted) ordering, due to the phenomenon of eigenvalue repulsion in random matrices.  
We consider three models for neutrino mass generation: Dirac, Majorana, and Majorana via the seesaw mechanism; differences in the predicted priors $\pi(\Sigma m_\nu)$ allow for the possibility of having indications about the physical origin of neutrino masses once sufficient experimental sensitivity is achieved.  
We present fitting functions for $\pi(\Sigma m_\nu)$, which provide a simple means for applying these priors to cosmological constraints on the neutrino masses or marginalizing over their impact on other cosmological parameters.  
\end{abstract}

\maketitle

\section{Introduction}\label{sec:Intro}

Measurements of the cosmic microwave background (CMB) \cite{Ade:2013zuv,Ade:2015xua} and large scale structure \cite{RiemerSorensen:2011fe,Palanque-Delabrouille:2014jca,Cuesta:2015iho} are currently being used to constrain -- and one day, hopefully to measure -- the spectrum of neutrino masses.
The two squared mass splittings are already known from terrestrial experiments \cite{Agashe:2014kda,King:2014nza}, and cosmological probes are sensitive to the sum of the three neutrino masses, $\Sigma m_\nu$.  
Therefore a combination of terrestrial and cosmological measurements can be used to completely determine the neutrino mass spectrum.  

When cosmological data is used to extract a limit on $\Sigma m_\nu$, the analysis is generally performed using Bayesian inference \cite{sivia2006data}.  
In this framework one is required to select a prior probability distribution $\pi(\Sigma m_\nu)$, which reflects one's {\it a priori} expectation for the sum of neutrino masses before the cosmological data is considered. 

The assumed prior can have a dramatic impact on the inferred posterior probability distributions, not only for the neutrino mass itself but also for other cosmological parameters due to degeneracies.  
Until recently, cosmological measurements did not have enough sensitivity for the prior to matter much, and for simplicity most analyses set $\Sigma m_\nu = 0$.  
However, the enhanced precision of the Planck CMB data \cite{Ade:2013zuv,Ade:2015xua}, meant that assuming $\Sigma m_\nu = 0$ would lead to a shift in the value of the Hubble constant (that is a noticeable fraction of its errors) as compared to assuming $\Sigma m_\nu \approx 0.06 \eV$, which is the minimum value allowed by the terrestrial experiments.  
Thereafter, most analyses allowed for non-zero neutrino mass by simply shifting the delta function prior from $\Sigma m_\nu = 0$ to $\Sigma m_\nu \approx 0.06 \eV$.  

We are now moving into the era where analyses need to account for the reality that we do not know the value of $\Sigma m_\nu$ by allowing it to vary in the fits.  
In choosing a prior for $\Sigma m_\nu$ we have more options than simply the flat and logarithmic distributions; since there are three neutrino masses, there are other options that seem closer to the fundamental parameters, and these options may lead to dramatic differences in the inferred parameter constraints.  
For example, the authors of \rref{Simpson:2017qvj} advocate a Gaussian (hyper)prior on the logarithm of the individual masses, and they find that given the mass splittings, the prior alone strongly favors minimal masses, the normally ordered spectrum over the inverted one, and the impact on other cosmological parameters that these preferences imply (cf.~\rref{Schwetz:2017fey}).

In \sref{sec:AdHoc} we review the importance of priors in establishing cosmological limits on $\Sigma m_\nu$.  
Even amongst commonly applied priors, the $95\%$ confidence upper limit can vary by as much as $20\%$.  
More extreme logarithmic priors can lead to qualitatively different conclusions.  
Therefore, understanding the origin of the prior has emerged as an essential ingredient of any cosmological analysis. 

Here we argue for a prior on the most fundamental parameters in the theory: the elements of the neutrino mass matrix $M_\nu$ that appears in the Lagrangian.  
This approach follows in the footsteps of earlier work on the ``anarchy hypothesis'' or ``anarchy principle'' in studies of the neutrino flavor problem \cite{Hall:1999sn} (see also Refs.~\cite{Haba:2000be,deGouvea:2003xe,Hall:2007zh,Hall:2007zj,Hall:2008km,deGouvea:2012ac,Lu:2014cla,Fortin:2016zyf,Babu:2016aro,Fortin:2017iiw}).  
By specifying a probability distribution for $M_\nu$, we derive a probability distribution over the sum of neutrino masses $\Sigma m_\nu$.  
Since the structure of the mass terms in the Lagrangian depends on whether the neutrinos are Dirac or Majorana particles and whether the Majorana mass arises from the seesaw mechanism, the distribution over $\Sigma m_\nu$ also depends on the physical origin for the neutrino masses.

On the one hand, this approach can be viewed as a phenomenological exercise.  
We are simply displacing the prior from $\Sigma m_\nu$ to $M_\nu$.  
When our work is viewed in this way, the interesting question is whether the flat prior on $\Sigma m_\nu$, which is typically assumed in cosmological studies, arises from reasonable, basis-independent priors on $M_\nu$; we will see that it does not.
(To motivate the flat prior, note that it is the Jeffreys prior for a noise-dominated measurement.)  

There is an even more ambitious interpretation of this approach. 
It is reasonable to treat the neutrino mass matrix as random if the fundamental theory admits many vacua across which the neutrino mass spectrum varies.  
Then the prior on $M_\nu$ reflects the uncertainty inherent in the underlying theory.
The landscape of string theory vacua is a concrete example \cite{Susskind:2003kw}, but of course string theory is not developed to a point where the probability distribution over $M_\nu$ can be calculated.  

\begin{figure}[t]
\includegraphics{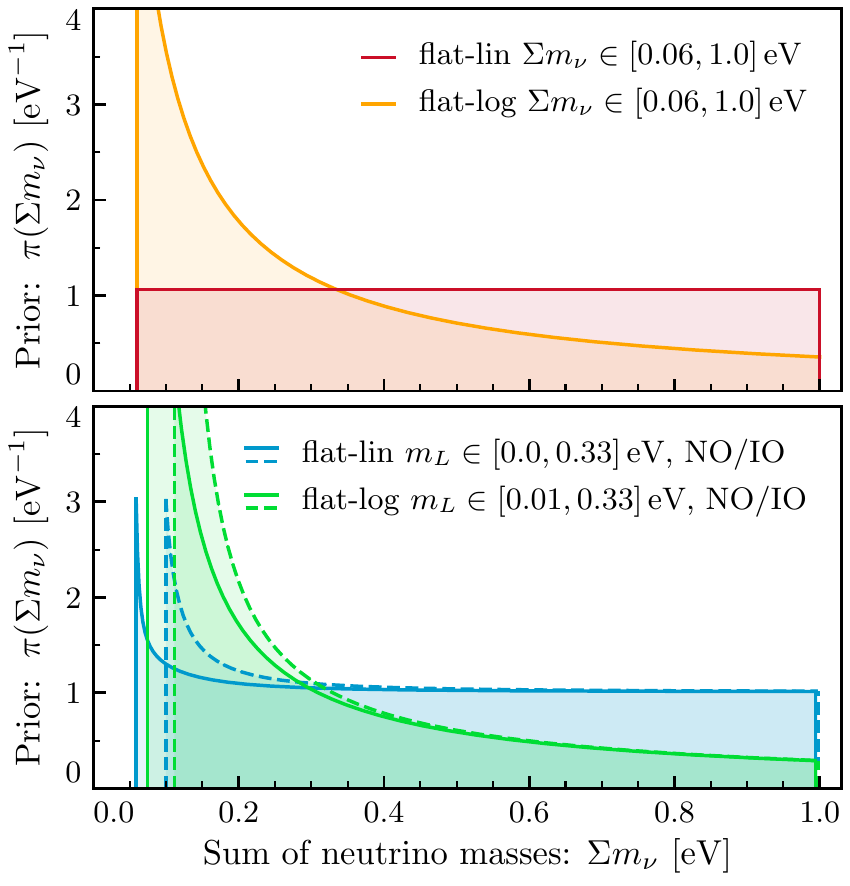} 
\caption{\label{fig:1}
Ad hoc choices for the prior probability on $\Sigma m_\nu$.}
\end{figure}

Nevertheless, if we understand that the random matrix description arises from some unknown fundamental theory, then we can restrict the form of the prior on $M_\nu$ with some reasonable assumptions about the character of that theory.  
Following \rref{Haba:2000be}, the primary assumption in our work is that the probability distribution over $M_\nu$ should be invariant under bi-unitary flavor transformations.  
In other words, if two matrices are related to one another by a change of basis, then they should be equally probable.  
Since this assumption is a specific implementation of the anarchy hypothesis \cite{Hall:1999sn}, we will refer to this principle as the basis-independent anarchy hypothesis (BAH).  
In this work we do not consider models in which the neutrino mass matrix encodes an underlying symmetry structure, although such models are also motivated from fundamental physical principles, and we do not expect the results of our analysis to apply to these models as well.  For an analysis of neutrino mass priors in hierarchal mass models we refer the reader to \rref{Bergstrom:2014owa}.  

We will see that the priors $\pi(\Sigma m_\nu)$ resulting from these simple assumptions are strongly informative compared with current data.  
This can either be viewed as undesirable for a prior or can be interpreted as a testable target for future measurements.
As a stark example, generically under the BAH, a detection of $\Sigma m_\nu = 0.1 \eV$ would rule out the seesaw mechanism at approximately $99.7\%$ CL significance.  

This remainder of this article is organized as follows.  
We illustrate in \sref{sec:AdHoc} the importance of neutrino mass priors with a simple example.  
\sref{sec:RandMat} introduces the reader to random matrices and explains how to calculate the probability distribution over $\Sigma m_\nu$.  
\sref{sec:Priors} discusses how a specific probability distribution over the neutrino mass matrix $M_\nu$ determines the probability distribution over $\Sigma m_\nu$.
The paper concludes in \sref{sec:Conc} where we summarize our results and discuss the possible implications for cosmological studies and limits on $\Sigma m_\nu$.  

\section{Importance of Neutrino Mass Priors}\label{sec:AdHoc}

Several recent studies \cite{Simpson:2017qvj,Schwetz:2017fey,Hannestad:2017ypp} (see also Refs.~\cite{DiValentino:2015sam,Hannestad:2016fog,Gerbino:2016ehw,Giusarma:2016phn,Gerbino:2016sgw,Couchot:2017pvz,Vagnozzi:2017ovm}) have investigated how much the prior on $\Sigma m_\nu$ affects limits on cosmological parameters and odds ratios for normal versus inverted ordering.  
In \fref{fig:1} we show several recently-discussed priors.  
The top panel shows both a flat in $\Sigma m_\nu$ (flat-linear) and a flat in 
$\log(\Sigma m_\nu)$ (flat-log) prior in the range $\Sigma m_\nu \in (0.06,1) \eV$.  
If the likelihood is Gaussian in $\Sigma m_\nu$ then the Jeffreys prior, which is calculated as the square root of the determinant of the Fisher information matrix, is flat-linear  $\Sigma m_\nu \in (0,\infty) \eV$ \cite{Jeffreys:1946,Kass:1996} (cf.~\cite{Simpson:2017qvj,Hannestad:2017ypp}).\footnote{The Jeffreys prior takes its shape from the constraining power of the data itself but is no more or less physically motivated than the other ad hoc priors.  For example  regions of parameter space where the data have no sensitivity are arbitrarily rejected by the prior.  Note also, Jeffreys' rule that a scale parameter should have a flat-log prior is not always in accord with the Jeffreys prior, especially for noise dominated measurements.}
The bottom panel shows priors on $\Sigma m_\nu$ that arise when the distribution over the lightest neutrino mass $m_L$ is assumed to be either flat-linear $m_L \in (0,0.33) \eV$ or flat-log $m_L \in (0.01,0.33) \eV$.
The other heavier neutrino masses are determined by fixing the squared mass splittings to their best-fit measured values, and the hierarchy is assumed to have either the normal or inverted ordering (NO or IO) where the larger splitting is between the heaviest or lightest neutrinos respectively; for additional details see \sref{sub:Sum_Split}.  
Note that in the flat-log case, the lower boundary is arbitrary since $m_L$, unlike $\Sigma m_\nu$, is not bounded by splitting data. 
We will return to this point below.  

Let us now illustrate how much these prior choices affect the inferred limit on $\Sigma m_\nu$ from a measurement $\Sigma \hat m_\nu$.  
We assume a Gaussian likelihood, $\Lcal({\Sigma \hat m_\nu} \, | \, \Sigma m_\nu) = \Ncal \, {\rm exp}[ - ({\Sigma \hat m_\nu} - \Sigma m_\nu)^2 / (2 \sigma^2) ]$ where the normalization factor $\Ncal$ ensures that $\int_{-\infty}^{\infty} \! \ud {\Sigma \hat m_\nu} \, \Lcal = 1$, and $\sigma$ models the experimental uncertainty.  
We consider the six priors $\pi(\Sigma m_\nu)$ that are shown in \fref{fig:1}.  
Assuming a measurement ${\Sigma \hat m_\nu} = 0$ and an uncertainty $\sigma = 0.3 \eV$, we calculate the posterior on $\Sigma m_\nu$ as $P(\Sigma m_\nu) = \Ecal^{-1} \Lcal(0 \eV \, | \, \Sigma m_\nu) \, \pi(\Sigma m_\nu)$ where the evidence $\Ecal$ ensures that $\int_{0}^{\infty} \! \ud \Sigma m_\nu \, P = 1$.  
The $95\%$ CL region is estimated to be 
\begin{align}\label{eq:adhoc_limits}
	\Sigma m_\nu / {\rm eV} < 0.61, \ 0.46, \ \ 0.59, \ 0.60, \ 0.42, \ 0.43
\end{align}
for the flat-linear prior on $\Sigma m_\nu$, the flat-log prior on $\Sigma m_\nu$, the flat-linear prior on $m_L$ with NO, flat-linear prior on $m_L$ with IO, flat-log prior on $m_L$ with NO, and flat-log prior on $m_L$ with IO.  
This exercise demonstrates that the $95\%$ CL upper limit on $\Sigma m_\nu$ varies by approximately $20\%$ for the family of priors shown in \fref{fig:1}.  

In fact, even larger variations in the upper limit on $\Sigma m_\nu$ can be obtained in principle.  
For the log prior on $m_L$ the lower boundary of $0.01 \eV$ is arbitrary, and it represents another ad hoc assumption -- namely, that the lightest neutrino's mass is not much smaller than the smaller mass splitting.  
However, suppose that we further decrease the lower boundary on $m_L$.  
Since most of the allowed prior volume is at small $m_L$, the prior probability increasingly favors the smallest possible value of $\Sigma m_\nu$ that is consistent with the splitting data.  
For example if we take the limit to $10^{-4} \eV$, the $95\%$ CL upper limit on $\Sigma m_\nu$ is strengthened by almost a factor of 2 from $0.42 \eV$ to $0.26 \eV$ for NO and from $0.43 \eV$ to $0.30 \eV$ for IO.
This is similar to the result found in \rref{Simpson:2017qvj}.  
There if one assumes that the individual masses are Gaussian distributed in the log, the normal ordering is also strongly preferred.  
The difference between our arbitrarily bounded flat case and the Gaussian case is that they then marginalize over the mean and width of the Gaussian so as to effectively place a lower bound that is comparable to the splittings.

In these more extreme examples, the prior assumptions qualitatively change the interpretation of the data.  
Without any physical principle to favor one prior over another, these changes simply represent ad hoc assumptions with no clear way to judge their merit or implications for neutrino physics.  
As such, the spread of values can be interpreted as a theoretical bias on upper limits for $\Sigma m_\nu$ that is bounded from below only by the minimal values allowed by the splitting data.

This situation will change as the cosmological data improve beyond the $10^{-1} \eV$ level and begin to detect $\Sigma m_\nu$.   
Compared to the prior choices in \fref{fig:1}, the data will become more informative than the prior, thereby reducing the $20\%$ scatter.  
On the other hand the more extreme case of a flat prior in $\log m_L$ with an arbitrarily small lower cutoff would continue to strongly favor a minimal $\Sigma m_\nu$.  
If the data measure otherwise, we would then reject the theory that produced this prior.  
However to reiterate, without a physical motivation for such a prior, it is difficult to assess the implications of that rejection for neutrino physics.  

Therefore in both the current and future contexts, it is important to make a stronger connection between the assumed priors and the neutrino physics that underly them.  
In particular we can ask the question whether any of the six ad hoc priors that were discussed above or the more extreme versions of the $\log m_L$ priors can be motivated by models of random neutrino mass matrices, which are more closely connected to models of neutrino mass at the Lagrangian level.

\section{Random Neutrino Mass Matrices}\label{sec:RandMat}

In this section we develop the formalism for lepton masses and random matrices.  
The reader will find \rref{RandMat:2009} to be a useful resource in the mathematical theory of random matrices.  
For applications to lepton matrices we mostly follow the original work on the anarchy hypothesis, and the equations in this section will be familiar from \rref{Haba:2000be}.  

\subsection{Charged Lepton and Neutrino Mass Matrices}\label{sub:Matrices}

Let the mass matrix for the three flavors of charged leptons (electron, muon, and tau) be denoted by $M_e$.  
In a general basis, $M_e$ need not be diagonal, but rather it is a 3-by-3 matrix with complex entries.
We can diagonalize $M_e$ with a singular value decomposition
\begin{align}\label{eq:Me_SVD}
	M_e = U_e \hat{M}_e V_e^{\dagger} \,,
\end{align}
where $U_e$ and $V_e$ are unitary matrices and $\hat{M}_e = {\rm diag}\bigl( m_{e1} , m_{e2} , m_{e3} \bigr)$ is the diagonal matrix of real and non-negative singular values, which correspond to the physical particle masses.  

The matrix $M_e$ is specified by $18$ real numbers, corresponding to the real and imaginary parts of each matrix element.  
Therefore each $M_e$ represents a point in the $18$-dimensional space $\Rbb^{18}$, and the measure in this space is defined by 
\begin{align}\label{eq:dMe_def}
	\ud^{18} M_e & = \Pi_{ij} \ \ud \, {\rm Re}[(M_e)_{ij}] \ \ud \, {\rm Im}[(M_e)_{ij}] 
	\per 
\end{align}
The factorization in \eref{eq:Me_SVD} specifies a change of coordinates from ${\rm Re}[M_e]$ and ${\rm Im}[M_e]$ to $\hat{M}_e$, $U_e$, and $V_e$.  
In terms of these new variables, the measure is written as 
\begin{align}\label{eq:dMe}
	\ud^{18} M_e = \Jcal_e(\hat{M}_e) \, \ud^3 \hat{M}_e \, \ud^9 U_e \, \ud^9 V_e / \ud^3 \varphi_e 
\end{align}
where 
\begin{align}\label{eq:Je}
	\Jcal_e(\hat{M}_e) & = m_{e1} m_{e2} m_{e3} \bigl( m_{e1}^2 - m_{e2}^2 \bigr)^2 
	\nn & \quad \times 
	\bigl( m_{e2}^2 - m_{e3}^2 \bigr)^2 \bigl( m_{e3}^2 - m_{e1}^2 \bigr)^2
\end{align}
is the determinant of the Jacobian of the transformation \pref{eq:Me_SVD}.  
Note that $\Jcal_e(\hat{M}_e)$ only depends upon the singular values of $M_e$.  
The measure over the space of singular values is written as $\ud^3 \hat{M}_e = \ud m_{e1} \, \ud m_{e2} \, \ud m_{e3}$.  
The measure over the space of unitary matrices, also known as the Haar measure, is denoted by $\ud^9 U_e$ for $U_e$ and $\ud^9 V_e$ for $V_e$.  
The unitary matrices can be parametrized by $3$ angles and $6$ phases, and an explicit expression for the Haar measure in these coordinates appears in \rref{Haba:2000be}.  
The SVD \pref{eq:Me_SVD} is not uniquely defined, but rather it remains invariant under a $\U{1}^3$ transformation, which takes $U_e \to U_e \Phi$ and $V_e \to V_e \Phi$ where $\Phi = {\rm diag}( e^{i \varphi_{e1}} , \, e^{i \varphi_{e2}} , \, e^{i \varphi_{e3}} )$.  
Consequently, the measure \pref{eq:dMe} is modded out by $\ud^3 \varphi_e = \ud \varphi_{e1} \, \ud \varphi_{e2} \, \ud \varphi_{e3}$.  

Let us next consider the mass matrix for the three flavors of neutral leptons (electron, muon, and tau neutrinos), which is denoted as $M_\nu$.  
Currently the origin of neutrino mass remains an open question.  
In general, any mass generation mechanism falls into one of two categories, which have implications for the structure of $M_\nu$.  
Either the neutrinos and antineutrinos are distinct Dirac particles, implying that $M_\nu$ is a 3-by-3 complex matrix, or the neutrinos and antineutrinos are identical Majorana particles, implying that $M_\nu$ is a 3-by-3 complex and symmetric matrix.  
The seesaw mechanism, which we discuss in \sref{sub:Seesaw}, is one specific scenario that gives rise to massive Majorana neutrinos.  
We treat the seesaw model separately, because (as we discuss in \sref{sub:Probability}) it is more natural to implement the probability distribution on the high-scale matrices rather than the low-energy neutrino matrix.

The symmetric mass matrix for Majorana neutrinos admits a Takagi decomposition, whereas the general mass matrix for Dirac neutrinos requires a singular value decomposition.  
Using these techniques, the neutrino mass matrix is factorized as 
\begin{align}\label{eq:Mnu_SVD}
	M_\nu = \begin{cases} 
	U_\nu \hat{M}_\nu V_\nu^\dagger & , \ \text{Dirac} \\
	U_\nu \hat{M}_\nu U_\nu^T & , \ \text{Majorana (e.g., seesaw)} 
	\end{cases} 
\end{align}
where $U_\nu$ and $V_\nu$ are unitary matrices and $\hat{M}_\nu = {\rm diag}\bigl( m_{\nu1} , m_{\nu2} , m_{\nu3} \bigr)$ is the diagonal matrix of real and non-negative singular values, which correspond to the physical neutrino masses.  

Having diagonalized both the charged and neutral lepton mass matrices (\ref{eq:Me_SVD}~and~\ref{eq:Mnu_SVD}), the lepton mixing matrix (PMNS matrix) is constructed as 
\begin{align}\label{eq:U_PMNS}
	U_{\PMNS} = U_e^{\dagger} U_{\nu} 
	\com
\end{align}
which quantifies the mismatch between the unitary transformations that diagonalize $M_e$ and $M_\nu$.  

The space of matrices $M_\nu$ is $18$-dimensional if the neutrinos are Dirac particles, but it is only $12$-dimensional if the neutrinos are Majorana particles, because the Majorana condition requires $M_\nu$ to be symmetric.  
Thus the measure over $M_\nu$ is defined by 
\begin{align}\label{eq:dM_nu}
	& \begin{cases}
	\ud^{18} M_\nu = \Pi_{ij} \ \ud \, {\rm Re}[(M_\nu)_{ij}] \ \ud \, {\rm Im}[(M_\nu)_{ij}] & , \ \mathrm{D} \\
	\ud^{12} M_\nu = \Pi_{i \leq j} \ \ud \, {\rm Re}[(M_\nu)_{ij}] \ \ud \, {\rm Im}[(M_\nu)_{ij}] & , \ \mathrm{M} 
	\end{cases} 
	\, . 
\end{align}
The decompositions in \eref{eq:Mnu_SVD} provide a change of coordinates, and the measure becomes 
\begin{align}\label{eq:dMnu}
	\begin{cases}
	\ud^{18} M_\nu = \Jcal_\nu(\hat{M}_\nu) \, \ud^3 \hat{M}_\nu \, \ud^9 U_\nu \, \ud^9 V_\nu / \ud^3 \varphi_\nu & , \ \mathrm{D} \\
	\ud^{12} M_\nu = \Jcal_\nu(\hat{M}_\nu) \, \ud^3 \hat{M}_\nu \, \ud^9 U_\nu & , \ \mathrm{M} 
	\end{cases}
\end{align}
where the Jacobian determinant is 
\begin{align}\label{eq:J_nu}
	& \Jcal_\nu(\hat{M}_\nu) = m_{\nu1} m_{\nu2} m_{\nu3} \\ 
	& \times 
	\begin{cases}
	\bigl( m_{\nu1}^2 - m_{\nu2}^2 \bigr)^2 \bigl( m_{\nu2}^2 - m_{\nu3}^2 \bigr)^2 \bigl( m_{\nu3}^2 - m_{\nu1}^2 \bigr)^2 & \! \! \! , \ \mathrm{D} \\
	\bigl| m_{\nu1}^2 - m_{\nu2}^2 \bigr| \bigl| m_{\nu2}^2 - m_{\nu3}^2 \bigr| \bigl| m_{\nu3}^2 - m_{\nu1}^2 \bigr| & \! \! \! , \ \mathrm{M} 
	\end{cases} \nonumber 
	\, , 
\end{align}
the measure over singular values is $\ud^3 \hat{M}_\nu = \ud m_{\nu1} \, \ud m_{\nu2} \, \ud m_{\nu3}$, and the measure over unitary matrices is defined in the same way as for $M_e$ [see the text below \eref{eq:Je}].  
The factors of $m_{\nu i}^2 - m_{\nu j}^2$ correspond to the phenomenon of eigenvalue repulsion, which probabilistically disfavors degeneracy; this effect is more pronounced in the Dirac model than the Majorana model.

\subsection{Majorana Masses from the Seesaw Mechanism}\label{sub:Seesaw}

One compelling scenario for the generation of a Majorana mass is the seesaw mechanism \cite{Minkowski:1977sc, Mohapatra:1979ia, GellMann:1980vs, Yanagida:1980xy, Mohapatra:1980yp, Schechter:1980gr}.  
The Type-I seesaw model contains a 3-by-3 complex matrix $M_D$ and a 3-by-3 complex, symmetric matrix $M_M$.  
In the Seesaw regime where the singular values of $M_M$ are much larger than the singular values of $M_D$, the neutrino mass matrix is well-approximated by 
\begin{align}\label{eq:M_nu}
	M_\nu = - M_D M_M^{-1} M_D^T 
	\per
\end{align}
Note that $M_\nu$ is a 3-by-3 complex, symmetric matrix.  
The measures $\ud^{18} M_D$ and $\ud^{12} M_M$ are defined in analogy with the Dirac and Majorana cases of $M_\nu$ \pref{eq:dM_nu}.  
Eigenvalue repulsion in $M_\nu$ is also strong in the seesaw case as demonstrated empirically in \rref{Haba:2000be}.  

\subsection{Probability Over Matrices}\label{sub:Probability}

Now we promote $M_e$, $M_\nu$, $M_D$, and $M_M$ to be random matrices.
The differential probability measure is written as 
\begin{align}\label{eq:dP_f}
	\ud \Pbb = \begin{cases} 
	f(M_e, M_\nu) \, \ud^{18} M_e \, \ud^{18} M_\nu & , \ \mathrm{D} \\ 
	f(M_e, M_\nu) \, \ud^{18} M_e \, \ud^{12} M_\nu & , \ \mathrm{M} \\
	F(M_e, M_D, M_M) \ \ud^{18} M_e \, \ud^{18} M_D \, \ud^{12} M_M & , \ \mathrm{S} 
	\end{cases} 
\end{align}
which define the Dirac (D), Majorana (M), and Seesaw (S) models, respectively.  
The probability densities, $f$ and $F$, are real-valued, positive, integrable, and normalized such that $\int \! \ud \Pbb =1$.  

Since the Seesaw model is also a model of Majorana neutrinos, it may seem redundant to distinguish these cases.  
There are various models in which high-scale lepton-number violation gives rise to a Majorana neutrino mass matrix, but since the Seesaw model is a particularly compelling and well-studied scenario, we consider this model separately.
If the Majorana mass does arise from the seesaw mechanism, then it is possible to derive $f(M_e, M_\nu)$ from $F(M_e, M_D, M_M)$. 
To make this connection concrete, first invert the Seesaw relation \pref{eq:M_nu} to determine the Majorana mass matrix as $M_M = - M_D^T M_\nu^{-1} M_D$, which defines a transformation from $M_M$ to $M_\nu$.  
Let the Jacobian of this transformation be denoted as $\Jcal_{M\nu}(M_D, M_\nu)$, such that $\ud^{18} M_D \, \ud^{12} M_M = \Jcal_{M\nu}(M_D,M_\nu) \ \ud^{18}M_D \, \ud^{12}M_\nu$.  
Marginalizing over the unobservable matrix $M_D$ yields 
\begin{align}\label{eq:F_to_f}
	f(M_e, M_\nu) & = \int \! \ud^{18} M_D \, \Jcal_{M\nu}(M_D, M_\nu)
	\\ & \qquad \times F(M_e, M_D, - M_D^T M_\nu^{-1} M_D) 
	\com \nonumber 
\end{align}
which is the same probability density that appears in the Majorana case of \eref{eq:dP_f}.  
In practice, we find that it is easier to study the Seesaw model by sampling from the $M_D$ and $M_M$ matrices directly rather than evaluating the Jacobian and integrals that appear in \eref{eq:F_to_f}.  

\subsection{Basis-Independent Anarchy Hypothesis}\label{sec:basis_invar}

The basis-independent anarchy hypothesis (BAH) provides guidance in selecting the probability densities $f$ and $F$ that appear in \eref{eq:dP_f}.  
In this section we clarify the meaning and implications of the BAH.  
Specifically, we show how the Lagrangian transforms under a basis-changing flavor transformation, and we require the probability distribution over matrices to be invariant under this transformation.
A reader who is more interested in the phenomenological implications may choose to skip this section.  

In writing the Lagrangian, it is convenient to work with the 2-component spinor notation \cite{Dreiner:2008tw}.  
The six charged leptons (electron, muon, tau, and anti-particles) are represented by the six fields $(e_L)_i$ and $(e_R)_i$ for $i = 1,2,3$, and the neutrinos are represented by the fields $(\nu_L)_i$ and $(\nu_R)_i$.  
The subscripts $L$ and $R$ indicate left- and right-chiral Weyl spinors, respectively.  
The masses and weak interactions of these fields are encoded in the Lagrangian\footnote{The mapping onto 4-component spinor notation is explained in \rref{Dreiner:2008tw}.  For instance, $\nu_L^\dagger \bar{\sigma}^\mu e_L W^+_\mu + \hc = \overline{\Psi}_\nu \gamma^\mu P_L \Psi_e W^+_\mu + \hc$ and $e_L^\dagger M_e e_R + \hc = \overline{\Psi}_e M_e \Psi_e$ and $\nu_L^T M_\nu \nu_L + \hc = \overline{\Psi_\nu^c} M_\nu \Psi_\nu$.  } 
\begin{align}\label{eq:L}
	\Lcal & \supset \frac{g}{\sqrt{2}} \nu_L^\dagger \bar{\sigma}^\mu e_L W^+_\mu  - e_L^\dagger M_e e_R + \Ocal_{\nu} + \hc 
\end{align}
where the vector field $W_\mu^+$ represents the two charged $W$-bosons, and where
\begin{align}
	\Ocal_\nu & = \begin{cases} 
	- \nu_L^\dagger M_\nu \nu_R & , \ \mathrm{D} \\
	- \frac{1}{2} \nu_L^T M_\nu \nu_L & , \ \mathrm{M} \\
	- \nu_L^\dagger M_D \nu_R - \frac{1}{2} \nu_R^T M_M \nu_R & , \ \mathrm{S} 
	\end{cases}
\end{align}
are the neutrino mass terms.  
In the Seesaw model, integrating out the heavy fields $\nu_R$ gives rise to a Majorana mass matrix $M_\nu$ for the light neutrinos, which is of the Seesaw form \pref{eq:M_nu}, $M_\nu = - M_D M_M^{-1} M_D^T$.  

Now consider a basis-changing flavor transformation.  
The fields in the new basis (primed) can be expressed as 
\begin{align}\label{eq:transformations}
	& 
	e_L^\prime = U_{e_L} e_L 
	\, , \ \ 
	\nu_L^\prime = U_{\nu_L} \nu_L 
	\, , \
	e_R^\prime = U_{e_R} e_R 
	\, , \
	\\ & \quad 
	\nu_R^\prime = U_{\nu_R} \nu_R 
	\, , \ \text{and} \ 
	W^{+ \prime}_\mu = W^+_\mu 
	\nonumber
\end{align}
where $U_{e_L}$, $U_{\nu_L}$, $U_{e_R}$, and $U_{\nu_R}$ are unitary matrices ($U U^\dag = 1$).  
The mass terms in \eref{eq:L} take the same form in the new basis provided that we identify the mass matrices in the new basis (primed) as 
\begin{align}\label{eq:matrix_transform}
	& M_e^\prime = U_{e_L} M_e U_{e_R}^\dagger 
	\, , \ \ 
	M_D^\prime = U_{\nu_L} M_D U_{\nu_R}^\dagger 
	\, , \ \ 
	\\ & \quad 
	M_M^\prime = U_{\nu_R}^\ast M_M U_{\nu_R}^\dagger 
	\, , \ \ 
	\nn & \ \text{and} \quad 
	M_\nu^\prime = \begin{cases} 
	U_{\nu_L} M_\nu U_{\nu_R}^\dagger & , \ \mathrm{D} \\
	U_{\nu_L} M_\nu U_{\nu_L}^T & , \ \mathrm{M} \\
	U_{\nu_L} M_\nu U_{\nu_L}^T & , \ \mathrm{S} 
	\end{cases}
	\nonumber \per
\end{align}
The weak interaction [first term in \eref{eq:L}] is not invariant in general, but instead it transforms to $(g / \sqrt{2}) (\nu_L^\prime)^\dagger \bar{\sigma}^\mu ( U_{\nu_L} U_{e_L}^\dagger ) e_L^\prime W_\mu^+$.  

The physical motivation for the anarchy hypothesis is that some unspecified high energy physics provides an ensemble of theories, i.e.\ vacua, across which the lepton masses vary.  
The principle of basis-independence further asserts that mass matrices related by a basis-changing flavor transformation should be equally probable.  
In other words, basis-independence restricts the allowed form of the probability densities $f(M_e, M_\nu)$ and $F(M_e,M_D,M_M)$ that we introduced in \eref{eq:dP_f}.  
In general there are various ways to implement basis independence:  
\setlist[description]{font=\normalfont}
\begin{description}[labelwidth=25pt, align=right,leftmargin=28pt]
	\item[BAH]  Require the probability densities to satisfy $f(M_e, M_\nu) = f(M_e, M_\nu^\prime)$ and $F(M_e, M_D, M_M) = F(M_e, M_D^\prime, M_M^\prime)$, which enforces basis-independence only on the neutrino mass matrices.  
	\item[B2]  Require the probability densities to satisfy $f(M_e, M_\nu) = f(M_e^\prime, M_\nu^\prime)$ and $F(M_e, M_D, M_M) = F(M_e^\prime, M_D^\prime, M_M^\prime)$ where the primed matrices are given by \eref{eq:matrix_transform} with $U_{e_L} = U_{\nu_L}$.  
	\item[B3]  Require the probability densities to satisfy $f(M_e, M_\nu) = f(M_e^\prime, M_\nu^\prime)$ and $F(M_e, M_D, M_M) = F(M_e^\prime, M_D^\prime, M_M^\prime)$ for arbitrary $U_{e_L}$ and $U_{\nu_L}$.  
\end{description}
The basis-independent anarchy hypothesis of \rref{Haba:2000be} is denoted by BAH, whereas B2 and B3 correspond to alternative strategies for enforcing basis independence.  

As argued in \rref{Hall:1999sn} the large lepton mixing angles and the comparable neutrino masses does not point to any underlying (flavor symmetry) structure in the neutrino sector.  
This motivated the authors of \rref{Hall:1999sn} to propose the anarchy hypothesis.  
In the charged lepton sector, on the other hand, the masses are very hierarchical ($m_e : m_\mu : m_\tau = 1 : 200 : 3500$), which suggests that a different physical mechanism is at play, {\it e.g.}\ the Froggatt-Nielsen mechanism.  
Hence, basis independence is imposed only on the neutrino mass matrix through the BAH, rather than on both lepton mass matrices as with either B2 or B3.  

The BAH requires that the probability densities are constructed from basis-invariant functions of $M \in \{ M_D, M_M, M_\nu \}$ that include the determinant $d = \abs{ \det{M} }$ and the traces $t_n = {\rm Tr}\bigl[ ( M^\dagger M )^n \bigr]$.  
Note that these functions only depend upon the singular values of the corresponding matrix.  
If we use these functions to build the probability densities, it follows that the distributions over angular variables, such as the phases and mixing angles in $U_{\PMNS}$, are simply given by the Haar measure.  
In this way, the assumption of basis-independence leads to a powerful prediction for the PMNS matrix parameters \cite{Haba:2000be}.  

Let us close this section by remarking upon the distinction between  BAH, {\rm B2}, and {\rm B3}.  
If the stochastic nature of the mass matrices arises from high energy physics above the weak scale, then the interactions of the leptons should respect $\SU{2}_L$ gauge invariance.  
This is implemented by combining $e_L$ and $\nu_L$ into the lepton doublet $L = (\nu_L \, , \, e_L)$ and requiring them to transform in the same way under flavor transformations.  
Hence in this context it is only reasonable to expect $f$ and $F$ to be invariant under flavor transformations that obey $U_{e_L} = U_{\nu_L}$, i.e. functions satisfying B2.  
The set of functions that satisfy B3 also satisfy BAH and B2.  
In fact BAH and B2 imply B3 but not all functions that satisfy B2 also satisfy BAH, for instance ${\rm Tr}\bigl[ M_e M_e^\dagger M_\nu M_\nu^\dagger \bigr]$.  
Upon factorizing $M_e$ and $M_\nu$ with \erefs{eq:Me_SVD}{eq:Mnu_SVD}, we can write this function as ${\rm Tr}\bigl[ \hat{M}_e^2 U_{\PMNS} \hat{M}_\nu^2 U_{\PMNS}^\dagger \bigr]$, which depends on both the singular values ($\hat{M}_e$ and $\hat{M}_\nu$) and the basis-invariant PMNS matrix \pref{eq:U_PMNS}.  
Thus if we were to enforce basis independence with {\rm B2} rather than the BAH, we would find that the probability density may depend directly on the PMNS matrix.   
However, we chose instead to enforce basis independence with the BAH, because the hierarchical nature of the charged lepton masses points to an underlying symmetry principle for their mass generation, as we have already discussed above.  

\subsection{Probability Densities after BAH}\label{sub:Densities}

Following Refs.~\cite{Hall:1999sn,Haba:2000be} let us now adopt the BAH assumption to write a simplified expression for the probability density $f(M_e, M_\nu)$ that appears in \eref{eq:dP_f}.  
In this section we focus on only the Dirac and Majorana models, since the corresponding expressions for the Seesaw model are more cumbersome and less illuminating.  
The BAH implies that the distribution over the neutrino mass matrix is independent from the distribution over the charged lepton matrix, and it follows that $f(M_e, M_\nu)$ can be written as a sum of factorizable components.  
We focus on the simplest case for which 
\begin{align}\label{eq:f_to_p}
	f(M_e, M_\nu) = p_e(M_e) \ p_\nu(\hat{M}_\nu)
\end{align}
where $p_\nu$ only depends on $M_\nu$ through the singular values $\hat{M}_\nu$.  
Then upon using the matrix measure decomposition in \eref{eq:dMnu} the probability measure \pref{eq:dP_f} becomes $\ud \Pbb = \ud \Pbb_e \, \ud \Pbb_\nu$ where $\ud \Pbb_e = p_e(M_e) \, \ud^{18} M_e$ and 
\begin{align}
	\ud \Pbb_\nu = 
	\begin{cases}
	p_\nu( \hat{M}_\nu) \, \Jcal_\nu(\hat{M}_\nu) \, \ud^3 \hat{M}_\nu \, \ud^9 U_\nu \ud^9 V_\nu / \ud^3 \varphi_\nu & , \ \mathrm{D} \\
	p_\nu( \hat{M}_\nu) \, \Jcal_\nu(\hat{M}_\nu) \, \ud^3 \hat{M}_\nu \, \ud^9 U_\nu & , \ \mathrm{M} 
	\end{cases}
	\per
\end{align}
The $\ud \Pbb$'s are separately normalized such that $\int \! \ud \Pbb = 1$.  
Noting that $\ud^{18} M_e \supset \ud^9 U_e$ and $\ud^9 U_e \, \ud^9 U_\nu = \ud^9 U_e \, \ud^9 U_\PMNS$, we see that the distribution over the PMNS angles and phases is simply given by the Haar measure $\ud^9 U_\PMNS$ \cite{Haba:2000be}.  
Since we are primarily interested in the distribution over masses, we integrate over the angular variables to obtain 
\begin{align}\label{eq:dP_nu}
	\ud \Pbb_\nu = 
	\begin{cases}
	p_\nu( \hat{M}_\nu) \, \Jcal_\nu(\hat{M}_\nu) \, \ud^3 \hat{M}_\nu \, \Vcal_2 & , \ \mathrm{D} \\
	p_\nu( \hat{M}_\nu) \, \Jcal_\nu(\hat{M}_\nu) \, \ud^3 \hat{M}_\nu \, \Vcal_1 & , \ \mathrm{M} 
	\end{cases}
\end{align}
where $\Vcal_1 = 2 \pi^6 / 3$ and $\Vcal_2 = \pi^9 / 3$ are the volumes of the compact angular spaces.  

\subsection{Summed Mass and Squared Splittings}\label{sub:Sum_Split}

The neutrino mass spectrum is not known, but rather the sum of the masses, $\Sigma m_\nu = m_{\nu1} + m_{\nu2} + m_{\nu3}$, has been constrained, and the squared mass splittings have been measured.  
Conventionally the mass splittings are defined as $\delta m^2 = m_{\nu2}^2 - m_{\nu1}^2$ and $\Delta m^2 = m_{\nu3}^2 - ( m_{\nu1}^2 + m_{\nu2}^2 )/2$ where the three mass eigenvalues are identified by the flavor content of the corresponding neutrino eigenstate (e.g., conventionally $m_{\nu3}$ is the mass of the neutrino with the smallest $\nu_e$ content).  
However, as discussed in \sref{sub:Probability}, we are interested in basis-independent probability measures over the space of neutrino mass matrices that lead to probability measures over the mass eigenvalues that are independent of the mixing angles (flavor composition).  
In other words, $m_{\nu1}$, $m_{\nu2}$, and $m_{\nu3}$ have the same statistics.  
Consequently, the data cannot be implemented as $\delta m^2 = m_{\nu2}^2 - m_{\nu1}^2 \sim 10^{-5} \eV^2$, for instance, since this expression singles out $\nu_2$ and $\nu_1$.  
In fact, the data tells us that the splitting between {\it any two} masses must be $10^{-5} \eV^2$; it could be $m_{\nu2}^2 - m_{\nu1}^2 \sim 10^{-5} \eV^2$, but it could also be $m_{\nu1}^2 - m_{\nu2}^2 \sim 10^{-5} \eV^2$ or $m_{\nu3}^2 - m_{\nu1}^2 \sim 10^{-5} \eV^2$ just as well.  

To identify a basis-independent parametrization of the spectrum, let us begin by defining $m_H$, $m_M$, and $m_L$ to be the masses of the heaviest, medium, and lightest neutrinos, respectively.
These masses are basis-independent functions of the spectrum.  
Next let us define 
\begin{subequations}\label{eq:123_to_SdD}
\begin{align}
	\Sigma m_\nu & = m_L + m_M + m_H \\ 
	\Delta m_{HM}^2 & = m_H^2 - m_M^2 \\ 
	\Delta m_{ML}^2 & = m_M^2 - m_L^2 
\end{align}
\end{subequations}
where $\Delta m_{HM}^2$ ($\Delta m_{ML}^2$) is the squared mass splitting between the two heavier (lighter) mass eigenstates.  
It is useful to define $h = {\rm sign}\bigl[ \Delta m_{HM}^2 - \Delta m_{ML}^2 \bigr]$ that lets us distinguish the two scenarios 
\begin{subequations}\label{eq:NO_IO_def}
\begin{align}
	{\rm NO}: & \quad h = +1 \ , \quad \Delta m_{HM}^2 > \Delta m_{ML}^2 \\ 
	{\rm IO}: & \quad  h = -1 \ , \quad \Delta m_{HM}^2 < \Delta m_{ML}^2
	\com 
\end{align}
\end{subequations}
which define normal ordering (smaller splitting on bottom) and inverted ordering (smaller splitting on top).  

To connect with measurements of the neutrino mass spectrum, it is useful to introduce the squared mass splittings $\delta m^2$ and $|\Delta m^2|$.  
Let us define, 
\begin{subequations}\label{eq:delta_defs}
\begin{align}
	\delta m^2 & = {\rm Min}\bigl[ \Delta m_{HM}^2 , \, \Delta m_{ML}^2 \bigr] \\ 
	|\Delta m^2| & = {\rm Max}\bigl[ \Delta m_{HM}^2 , \, \Delta m_{ML}^2 \bigr] + \delta m^2 / 2
\end{align}
\end{subequations}
such that $\delta m^2$ is the smaller of the two squared mass splittings, and $|\Delta m^2|$ is the larger of the two squared mass splittings plus half the smaller splitting.  
For example, consider a spectrum for which $m_{\nu1} < m_{\nu2} < m_{\nu3}$ and the smaller splitting is on the bottom, $m_{\nu2}^2 - m_{\nu1}^2 < m_{\nu3}^2 - m_{\nu2}^2$.  
In this case, we can write $\delta m^2 = m_{\nu2}^2 - m_{\nu1}^2$ and $|\Delta m^2| = m_{\nu3}^2 - (m_{\nu1}^2 + m_{\nu2}^2) / 2$.  
Alternatively, consider a spectrum $m_{\nu3} < m_{\nu1} < m_{\nu2}$ with the smaller splitting on the top, $m_{\nu2}^2 - m_{\nu1}^2 < m_{\nu1}^2 - m_{\nu3}^2$. 
Then, $\delta m^2 = m_{\nu2}^2 - m_{\nu1}^2$ and $|\Delta m^2| = \bigl| m_{\nu3}^2 - (m_{\nu1}^2 + m_{\nu2}^2) / 2 \bigr|$.  
However, as we discussed at the start of this section, we do not let these relations with $(m_{\nu1}, m_{\nu2}, m_{\nu3})$ define the squared mass splittings, because they are not basis-independent.  

Observations of neutrino flavor oscillations (see \cite{Agashe:2014kda} for a review) have furnished measurements of the squared mass splittings $\delta m^2 = ( 7.37 \pm 0.17) \times 10^{-5} \eV^2$ and $|\Delta m^2| = ( 2.50 \pm 0.04 ) \times 10^{-3} \eV^2$ (NO) and $|\Delta m^2| = ( 2.46 \pm 0.04 ) \times 10^{-3} \eV^2$ (IO) \cite{Capozzi:2016rtj}.  
The data admits two scenarios for mass ordering \pref{eq:NO_IO_def} depending on whether the smaller splitting is on the bottom (NO) or on the top (IO).  
Let $P_{\rm data}( \Delta m_{HM}^2, \Delta m_{ML}^2 )$ denote a probability density over the squared mass splittings that encodes these measurements.  
Using \eref{eq:delta_defs} the mass splitting measurements imply that $P_{\rm data}$ is bi-modal with one peak at $\Delta m_{HM}^2 \approx |\Delta m^2|$ and $\Delta m_{ML}^2 \approx \delta m^2$ for $h = +1$ (NO), and a second peak at $\Delta m_{HM}^2 \approx \delta m^2$ and $\Delta m_{ML}^2 \approx |\Delta m^2|$ for $h = -1$ (IO).  

Equation~(\ref{eq:123_to_SdD}) defines a transformation from $( m_{\nu1}, m_{\nu2}, m_{\nu3} )$ to $( \Sigma m_\nu, \Delta m_{HM}^2, \Delta m_{ML}^2 )$.  
Let the corresponding Jacobian determinant be denoted as $\Jcal_{123}(\Sigma m_\nu, \Delta m_{HM}^2, \Delta m_{ML}^2)$ such that $\ud m_{\nu1} \, \ud m_{\nu2} \, \ud m_{\nu3} = \Jcal_{123} \, \ud \Sigma m_\nu \, \ud \Delta m_{HM}^2 \, \ud \Delta m_{ML}^2$.  
In terms of these new variables, the differential probability measure \pref{eq:dP_nu} becomes 
\begin{align}\label{eq:dP_SdD}
	\ud \Pbb_\nu & = P(\Sigma m_\nu, \Delta m_{HM}^2, \Delta m_{ML}^2 ) \\
	& \qquad \ud \Sigma m_\nu \, \ud \Delta m_{H}^2 \, \ud \Delta m_{ML}^2 \nonumber
\end{align}
where the probability density is given by 
\begin{align}\label{eq:P_SdD}
	& P(\Sigma m_\nu, \Delta m_{HM}^2, \Delta m_{ML}^2 ) \\
	& \ = p_\nu(\hat{M}_\nu ) \, \Jcal_\nu(\hat{M}_\nu) \, \Jcal_{123}(\Sigma m_\nu, \Delta m_{HM}^2, \Delta m_{ML}^2)
	\com \nonumber
\end{align}
and it is normalized such that $\int \! \ud \Pbb_\nu = 1$.  

Since the transformation from $(m_{\nu1}, m_{\nu2}, m_{\nu3})$ into $(m_H, m_M, m_L)$ involves minimization and maximization functions, it is non-analytic when any of the two masses are equal.  
This makes it difficult to calculate the Jacobian across the full parameter space.  
However, the problem simplifies significantly for probability densities $p_\nu(\hat{M}_\nu)$ that respect basis-independence.  
For a given triplet $(m_L, m_M, m_H)$ there are six possible ways to identify $(m_{\nu1}, m_{\nu2}, m_{\nu3})$.  
Thanks to the basis-independence of the prior, each of these parameter combinations must be equally probable.  
Thus, the full parameter space $(m_{\nu1}, m_{\nu2}, m_{\nu3})$ can be subdivided into six ``copies.''  
Without loss of generality, we can work in the region where $m_{\nu1} < m_{\nu2} < m_{\nu3}$, which lets us identify $m_L = m_{\nu1}$, $m_M = m_{\nu2}$, and $m_H = m_{\nu3}$.  
Working in this wedge of parameter space, the probability density is multiplied by $6$ to account for a sum over the copies.  

With the procedure described above it is straightforward to calculate $\Jcal_{123}$, but the expression is unwieldy, and we will not reproduce it here.  
However, it is important to understand how $\Jcal_{123}$ scales with $\Sigma m_\nu$.  
Note that $\Jcal_{123}$ has mass dimension $-2$, and one can verify that it scales as $\Jcal_{123} \sim (\Sigma m_\nu)^{-2}$ in the regime where $\Sigma m_\nu \gg \sqrt{|\delta m^2|}, \sqrt{|\Delta m^2|}$.  
For comparison, we determine the scaling of $\Jcal_\nu(\hat{M}_\nu)$ by inspecting \erefs{eq:J_nu}{eq:123_to_SdD}.  
It the regime where the splittings are small, we have 
\begin{align}\label{eq:Jnu_SdD}
	\Jcal_\nu \sim 
	\begin{cases}
	(\Sigma m_\nu)^3 \, (\delta m^2)^2 \, |\Delta m^2|^4  & , \ \mathrm{D} \\
	(\Sigma m_\nu)^3 \, (\delta m^2) \, |\Delta m^2|^2  & , \ \mathrm{M} 
	\end{cases}
	\per 
\end{align}
The factors of $\delta m^2$ and $|\Delta m^2|$ correspond to the eigenvalue repulsion that we discussed below \eref{eq:J_nu}.  
Then, the probability density in \eref{eq:P_SdD} goes as $P \sim p_\nu(\hat{M}_\nu) \times (\Sigma m_\nu)^1$, which is linearly rising in $\Sigma m_\nu$ if $p_\nu$ is flat.  
We will make use of this fact in the following sections.  

Equation~(\ref{eq:P_SdD}) gives the joint probability density over the summed neutrino mass and the squared mass splittings.  
We are also interested in the conditional probability over $\Sigma m_\nu$ after requiring the squared mass splittings to match their measured values and also requiring the hierarchy to be either normal or inverted.  
Thus we calculate 
\begin{widetext}
\begin{align}\label{eq:P_of_Sm_fix_mu}
	\pi(\Sigma m_\nu | h) 
	= \frac{6}{ {\cal E}_{\text{NO}}+{\cal E}_{\text{IO}}} \int & \ud \Delta m_{HM}^2 \, \ud \Delta m_{ML}^2 \ 
	P(\Sigma m_\nu, \Delta m_{HM}^2, \Delta m_{ML}^2) 
	\nn & \ \times
	P_{\rm data}( \Delta m_{HM}^2, \Delta m_{ML}^2 ) 
	\times \begin{cases} 
	\Theta(\Delta m_{ML}^2 < \Delta m_{HM}^2) & , \quad h = +1 \ \text{(NO)} \\ 
	\Theta(\Delta m_{HM}^2 < \Delta m_{ML}^2) & , \quad h = +1 \ \text{(IO)} 
	\end{cases}
	\per 
\end{align}
\end{widetext}
The bimodal probability distribution $P_{\rm data}$, which is defined below \eref{eq:delta_defs}, implements the mass splitting measurements.  
The step function, which satisfies $\Theta(x) = 1$ for $x \geq 0$ and $\Theta(x) = 0$ for $x < 0$, lets us condition on the hierarchy.  
The normalization factor ${\cal E}_{\text{NO}}+{\cal E}_{\text{IO}}$ is the sum of the evidences for the NO and IO cases separately so that $\sum_{h=\pm1} \int_{0}^{\infty} \ud \Sigma m_\nu \, \pi(\Sigma m_\nu | h) = 1$.  

\section{Basis-Independent Anarchy Priors on $\Sigma m_\nu$}\label{sec:Priors}

In the context of random mass matrices, the neutrino mass prior $\pi(\Sigma m_\nu)$ is precisely the marginalized probability from \pref{eq:P_of_Sm_fix_mu}.  
How does the BAH guide us in choosing a prior?   
We first show in \sref{sec:AdHocRevisited}, that the common ad hoc priors on $\Sigma m_\nu$ discussed in \sref{sec:AdHoc} appear highly contrived in the BAH context.
Arguably the three most natural choices for probability distributions over $M_\nu$ are flat, Gaussian, and logarithmic distributions over each matrix element.  
Interestingly the logarithmic distribution is forbidden by the BAH.  
The Gaussian distribution has a similar behavior to the flat distribution (see Sec.~\ref{app:Stability}), and so we focus on the latter in \sref{sec:Lin_Prior}.  

\subsection{Relationship to Ad Hoc Priors}\label{sec:AdHocRevisited}

In \sref{sec:AdHoc} we looked at a few ad hoc priors on $\Sigma m_\nu$, such as flat-linear and flat-log, that are commonly used for cosmological data analyses.  
In this section we discuss whether these priors can arise in the framework of random neutrino matrices subject to the BAH condition.  

In constructing the distribution $\pi(\Sigma m_\nu \, | \, h)$, the only freedom is to specify the probability density $p_\nu(\hat{M}_\nu)$ that appears in \eref{eq:P_SdD}.  
Recall that the requirement of basis-independence (BAH) means that we must construct $p_\nu(\hat{M}_\nu)$ from functions of the neutrino mass matrix $M_\nu$ that are left invariant under the basis-changing flavor transformation; see \sref{sec:basis_invar}.
This severely constraints the functional forms that we can use; examples of viable functions include the determinant, $d = |\det M_\nu| = m_{\nu1} m_{\nu2} m_{\nu3}$, and the traces, $t_n = {\rm Tr}[ \bigl( M_\nu^\dagger M_\nu \bigr)^n ] = (m_{\nu1}^2)^n + (m_{\nu2}^2)^n + (m_{\nu3}^2)^n$ for positive, integer $n$. 

Cosmological studies typically set a flat-linear prior on $\Sigma m_\nu$, and we can ask whether this choice is reasonable from the context of random neutrino matrices.  
Note that taking $p_{\nu}(\hat{M}_\nu) \propto \bigl[ \Jcal_\nu(\hat{M}_\nu) \bigr]^{-1}$ in \eref{eq:dP_nu} results in $\ud \Pbb \propto \ud m_1 \, \ud m_2 \, \ud m_3$, which is flat in each of the three neutrino masses.  
To obtain a distribution that is flat in $\Sigma m_\nu$ we should consider \erefs{eq:dP_SdD}{eq:P_SdD}.  
Recall that $\Jcal_{123} \sim (\Sigma m_\nu)^{-2}$ for $\Sigma m_\nu \gg \Delta m_{HM}^2 , \Delta m_{ML}^2$, as we discussed above \eref{eq:Jnu_SdD}.  
Thus, we should take $p_{\nu}(\hat{M}_\nu) \propto \bigl[ \Jcal_\nu(\hat{M}_\nu) \bigr]^{-1} ( m_1 m_2 m_3 )^{2/3}$, because $m_1 \sim m_2 \sim m_3 \sim \Sigma m_\nu/3$ in the degenerate regime.  
This results in a prior that is flat in $\Sigma m_\nu$ for $\Sigma m_\nu \gg \Delta m_{HM}^2 , \Delta m_{ML}^2$.  

To understand whether $p_{\nu}(\hat{M}_\nu) \propto \bigl[ \Jcal_\nu(\hat{M}_\nu) \bigr]^{-1} ( {\rm det} \, \hat{M}_\nu )^{2/3}$ is a reasonable prior, let us express it in terms of the matrix invariants.  
The Jacobian factor $\Jcal_\nu(\hat{M}_\nu)$ appears in \eref{eq:J_nu} where it is expressed as a function of the singular values $m_{\nu i}$. 
Writing $\Jcal_\nu$ instead in terms of the matrix invariants, $d$ and $t_n$ defined above, it takes the form $\Jcal_\nu = \Ccal d$ for Dirac neutrinos and $\Ccal^{1/2} d$ for Majorana neutrinos where $\Ccal = \bigl( -27 d^4 + 5 d^2 t_1^3 - \frac{1}{4} t_1^6 - 9 d^2 t_1 t_2 + t_1^4 t_2 - \frac{5}{4} t_1^2 t_2^2 + \frac{1}{2} t_2^3 \bigr)$. 
Thus taking $p_\nu \propto [\Jcal_\nu]^{-1} \, d^{2/3}$ leads to a distribution that is asymptotically flat in $\Sigma m_\nu$, but it gives $p_\nu$ a highly non-trivial dependence on $M_\nu$.
Similar arguments hold for all of the six ad hoc priors that were discussed in \sref{sec:AdHoc}. 

\subsection{Log Prior on Matrix Elements}\label{sec:Log_Prior}

It is customary to use a logarithmic prior for dimensionful ``scale'' parameters.  
Therefore it is natural to consider imposing such a prior for the neutrino mass matrix.  
In order to place a logarithmic prior on the real and imaginary parts of each matrix element, we should choose the  probability density $f(M_e, M_\nu)$ from \eref{eq:dP_f} to be proportional to $[ \Pi_{ij} ({\rm Re} \, M_{\nu})_{ij} ({\rm Im} \, M_{\nu})_{ij}]^{-1}$ with some cutoffs at small and large mass to ensure integrability.  
However, this product of matrix elements cannot be expressed in terms of the basis-independent matrix invariants, $d$ and $t_n$ from \sref{sec:AdHocRevisited}.  
Consequently a logarithmic prior on the matrix elements is not consistent with the BAH.  
This is an example of how theory-motivated assumptions like the BAH provide guidance in choosing a physically-motivated prior.  

\subsection{Flat Prior on Matrix Elements}\label{sec:Lin_Prior}

In this section we introduce a probability density that will be the focus of our extensive numerical study in the following sections.  
Consider the probability density that is flat in the real and imaginary parts of each element of the neutrino mass matrix $M_\nu$.  
In terms of the probability density $p_\nu(\hat{M}_\nu)$ from \erefs{eq:dP_f}{eq:f_to_p}, this distribution is expressed as 
\begin{align}\label{eq:lin_prior}
	p_\nu(\hat{M}_\nu \, | \, \mu_\nu) = \Ncal \, \Theta\bigl( \mu_\nu^2 - {\rm Tr}\bigl[ \hat{M}_\nu^\dagger \hat{M}_\nu \bigr] \bigr)
\end{align}
for the Majorana and Dirac models.  
Here $\Theta(x)$ is the unit step that evaluates to $\Theta = 1$ for $x \geq 0$ and $\Theta = 0$ otherwise.  
The parameter $\mu_\nu$ controls the neutrino energy scale.  
In the Seesaw model the probability density $F(M_e,M_D,M_M)$ from \eref{eq:dP_f} is chosen to be 
\begin{align}\label{eq:lin_priorSeesaw}
	& F( M_e, M_D, M_M \, | \, \mu_D, \mu_M ) \\
	& \quad = \Ncal \, 
	\Theta\bigl( \mu_D^2 - {\rm Tr}\bigl[ M_D^\dagger M_D \bigr] \bigr) 
	\Theta\bigl( \mu_M^2 - {\rm Tr}\bigl[ M_M^\dagger M_M \bigr] \bigr) 
	\nonumber \per
\end{align}
Since we are only interested in the spectrum of light neutrinos, which is given by the Seesaw relation \pref{eq:M_nu}, the relevant observables will only depend upon the effective neutrino scale
\begin{align}\label{eq:Seesaw_scale}
	\mu_\nu = \frac{\mu_D^2}{\mu_M} 
	\per
\end{align}
For the Dirac and Majorona models, notice that the step function imposes a hard upper bound on the neutrino mass spectrum, $m_\nu \leq \mu_\nu$ or $\Sigma m_\nu \leq \sqrt{3} \mu_\nu$, but there is no such bound for the Seesaw model.  

The normalization factors $\Ncal$ are obtained by imposing $\int \ud \Pbb = 1$ over the space of random matrices.  
For the Dirac model $\int \ud \Pbb$ equals $\Ncal$ times the volume of the $18$-dimensional ball\footnote{The volume of the $n$-dimensional unit ball is $\pi^{n/2} / \Gamma(n/2+1)$.} with radius $\mu_\nu$; for the Majorana model it equals $\Ncal$ times the volume of a $12$-dimensional ellipsoid; and for the Seesaw model $\int \ud \Pbb$ is the product of these two.  
Solving for the normalization factor gives:
\begin{align}\label{eq:Ncal}
	\Ncal = \begin{cases}
	\frac{1}{3} \Gamma(10) \, \Vcal_2^{-1} \,\, \mu_\nu^{-18}  & , \ \mathrm{D} \\
	\frac{16}{3} \Gamma(7) \, \Vcal_1^{-1} \,\, \mu_\nu^{-12} & , \ \mathrm{M} \\ 
	\frac{16}{9} \Gamma(7) \Gamma(10) \, \Vcal_1^{-1} \Vcal_2^{-1} \, \mu_M^{-12}\mu_D^{-18} & , \ \mathrm{S} 
	\end{cases} 
	\com
\end{align}
where $\Gamma(z)$ denotes the Gamma function, and where $\Vcal_1$ and $\Vcal_2$ were given below \eref{eq:dP_nu}.  
The high power of $\mu_\nu$ appears to compensate the mass dimensions of $\ud^{12} M_\nu$, $\ud^{18} M_\nu$, and $\ud^{18} M_D \ud^{12} M_M$ in the Majorana, Dirac, and Seesaw models, respectively; see \eref{eq:dP_f}. 

\subsection{Methodology}\label{sec:Analysis}

We have performed an extensive semi-analytical and numerical analysis of the probability distributions appearing in \erefs{eq:lin_prior}{eq:lin_priorSeesaw}.  
In the remainder of this section we describe the numerical techniques, and in the following section we describe the results for $\Sigma m_\nu$.  

A semi-analytical analysis is computationally feasible for the (M) and (D) models.  
Recall that the probability distribution over the neutrino mass sum and squared splittings was given by \eref{eq:P_SdD} where $p_\nu(\hat{M}_\nu)$ is given by \eref{eq:lin_prior}.  
Upon fixing the mass splittings to their best-fit measured values, the distribution over the neutrino mass sum is given by \eref{eq:P_of_Sm_fix_mu} for both NO and IO cases.  
These integrals are evaluated numerically with Mathematica.   

In addition we perform a fully numerical analysis for all three models: (M), (D), and (S).  
Here we implement the mass splitting measurements as a Gaussian likelihood.  
To perform the numerical integrals we Markov Chain Monte Carlo (MCMC) sample the full space of random matrices by means of the {\rm emcee} algorithm \cite{ForemanMackey:2012ig}.  
We analyze the resulting distributions with the {\rm GetDist} package \cite{getDist}.
To compute the overall normalization of the probability densities we use the algorithm described in \cite{Heavens:2017afc} and direct nested sampling integration with {\rm multinest} \cite{Feroz:2008xx,Buchner:2014nha}.
We shall hereafter refer to the normalization of the probability densities as the evidence, $\mathcal{E}$. 

For the (M) and (D) models the semi-analytical analysis provides a test of the accuracy of the MCMC integration, which is the only technique that we use to study the (S) model.  
Overall, we find excellent agreement for the (M) and (D) models (see \fref{fig:7} below), and these tests further validate the pipeline that we apply to the (S) model.  
In all three models the precision of the results is also estimated with several split-tests on the MCMC samples. In particular we require stability of the results to: residual burn-in samples by discarding relevant fractions of the initial samples; subdivision of the samples in uncorrelated and correlated sub-samples; and convergence of the results by discarding relevant fractions of the last samples. 

The difference between the evidence calculations for the semi-analytical and numerical cases
is  $\Delta \log_{10} \mathcal{E} = 0.2 \,(0.2)$ in the (M) model NO (IO) and $\Delta \log_{10} \mathcal{E} = 0.003\,(0.3)$ in the (D) case for NO (IO).  In these cases, the semi-analytical calculation is the more accurate and we use these in our results below.
This should be compared with the numerical error reported by multinest that reads $\Delta \log_{10} \mathcal{E} = 0.1\,(0.1)$ in the (M) case and $\Delta \log_{10} \mathcal{E} = 0.01\,(0.4)$ in the (D) case.
In the (S) case, where the semi-analytical results are not available, the multinest estimate of the evidence error $\Delta \log_{10} \mathcal{E} \sim 0.2\, (0.3)$ and the nested sampling and importance nested sampling estimates are within this error estimates. 

\begin{figure*}[t]
\begin{center}
\includegraphics{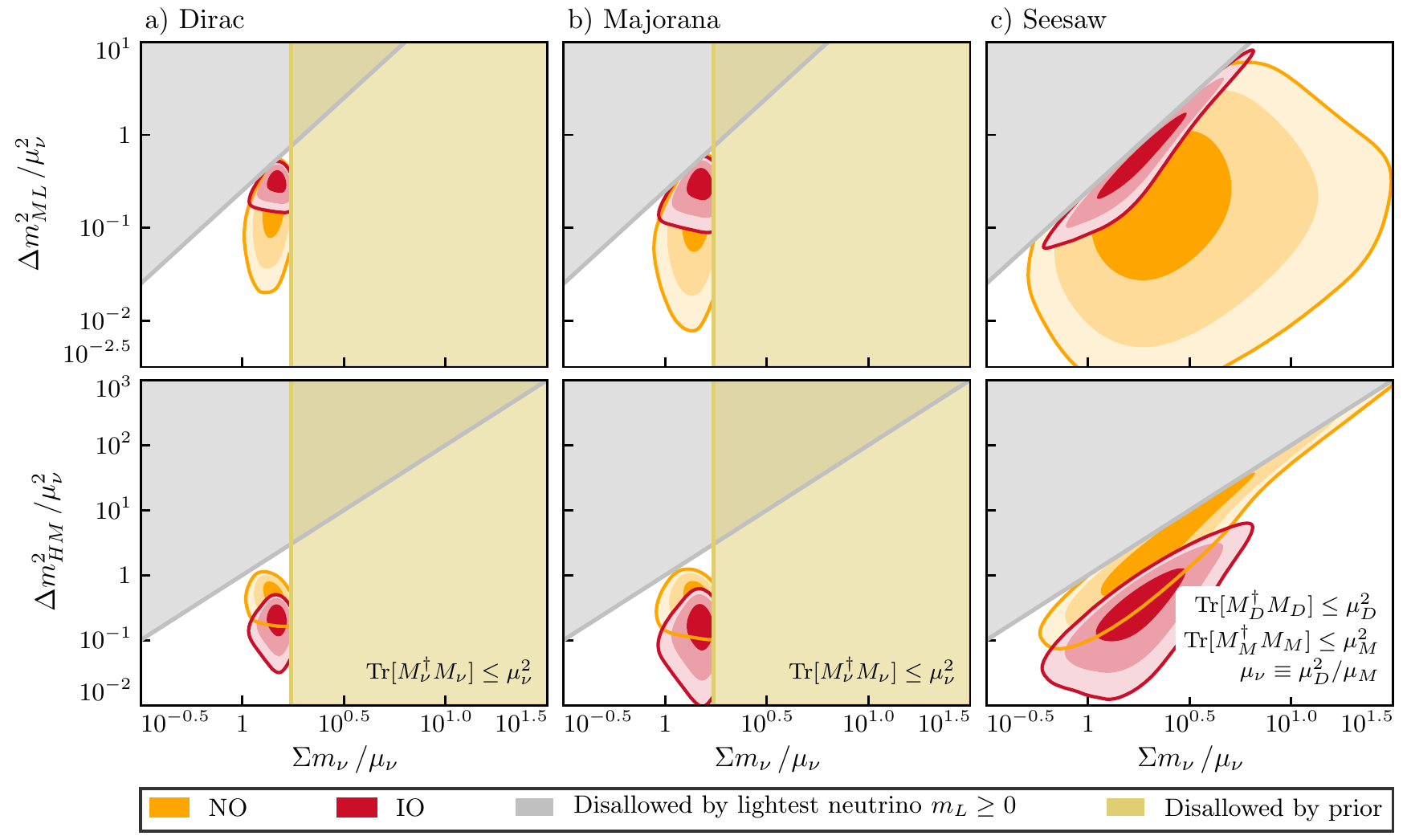} 
\caption{\label{fig:2}
The probability distribution over the mass sum, $\Sigma m_\nu = m_L + m_M + m_H$, and one of the squared mass splittings, $\Delta m_{HM}^2 = m_H^2 - m_M^2$ and $\Delta m_{ML}^2 = m_M^2 - m_L^2$, after marginalizing over the other splitting.  
The three panels correspond to different models for the origin of neutrino mass:  {\it Left}, we sample the Dirac neutrino mass matrix $M_\nu$ with a flat distribution on its matrix elements and cutoff ${\rm Tr}[M_\nu^\dagger M_\nu] \leq \mu_\nu^2$ \pref{eq:lin_prior}; {\it Middle}, we sample the (symmetric) Majorana neutrino mass matrix $M_\nu$ with a flat distribution on its matrix elements and cutoff ${\rm Tr}[M_\nu^\dagger M_\nu] \leq \mu_\nu^2$ \pref{eq:lin_prior}; {\it Right}, we sample the high-scale Dirac and Majorana mass matrices, $M_D$ and $M_M$, with a flat distribution on their matrix elements and separate cutoffs $\mu_D$ and $\mu_M$ \pref{eq:lin_priorSeesaw}, and then we use the seesaw relation \pref{eq:M_nu} to calculate the neutrino mass matrix, which only depends on the ratio $\mu_\nu = \mu_D^2 / \mu_M$.  
The orange (red) region indicates models with normal (inverted) mass ordering for which $\Delta m_{HM}^2 > \Delta m_{ML}^2$ ($\Delta m_{HM}^2 < \Delta m_{ML}^2$). 
Requiring $m_L > 0$ forbids the parameter space where $\Sigma m_\nu < \sqrt{ \Delta m_{HM}^2 + \Delta m_{ML}^2 } + \sqrt{ \Delta m_{ML}^2 }$, which corresponds to the upper-left gray triangle.  
Additionally the prior \pref{eq:lin_prior} forbids $\Sigma m_\nu > \sqrt{3} \mu_\nu$ in the Dirac and Majorana cases, which corresponds to the yellow rectangle in the left and middle panels.
The darker and lighter shades correspond respectively to the $68\%$ C.L., $95\%$ C.L. and $99.7\%$ C.L. regions.  
}
\end{center}
\end{figure*}

\begin{figure*}[t]
\begin{center}
\includegraphics{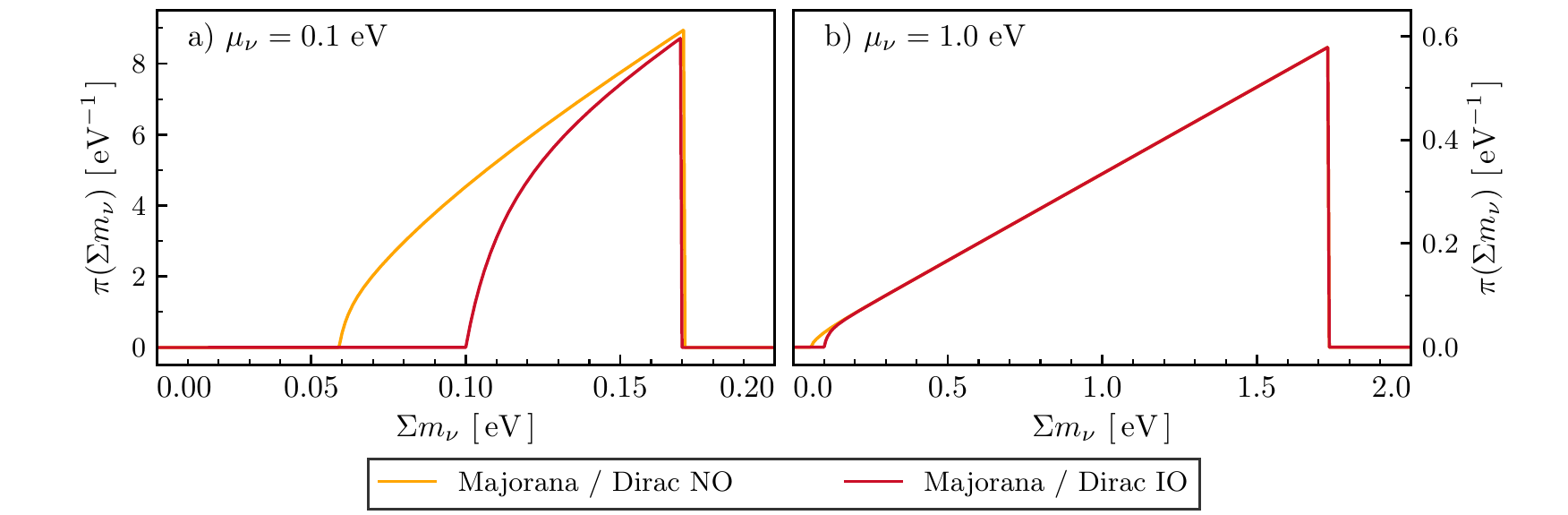} 
\caption{\label{fig:3}
The probability distribution over the sum of neutrino masses after the squared mass splittings are fixed to their best-fit measured values, and the neutrino mass scale is fixed to be $\mu_\nu = 0.1 \eV$ ($1.0 \eV$) in the left (right) panel.  
The shape of the probability distributions are identical for both the Majorana and Dirac models, but the shapes are distinguished for normal and inverted ordering.  
For a fixed $\mu_\nu$ the most probable value is $\Sigma m_\nu = \sqrt{3} \mu_\nu$, and increasing $\mu_\nu$ moves the peak to larger $\Sigma m_\nu$, but at the same time it reduces the overall probability of obtaining the observed squared mass splittings (not shown on the figure); i.e., the evidence decreases rapidly with increasing $\mu_\nu$ (see discussion in the text).
}
\end{center}
\end{figure*}

\subsection{Implications for $\Sigma m_\nu$}\label{sub:Implications}

Let us now discuss the numerical results.  
We first investigate the probability distributions over the summed neutrino masses and the squared mass splittings that are predicted from a flat-linear distribution on individual matrix elements (\ref{eq:lin_prior}~and~\ref{eq:lin_priorSeesaw}).  
At this point we do not fix the squared mass splittings to equal their measured values, but we explore the parameter space more broadly.  

Figure~\ref{fig:2} shows the joint distribution of the sum of the neutrino masses against the distribution of one squared mass splitting after marginalizing over the other squared mass splitting for both the NO and IO cases from the numerical analysis.  
Masses are expressed in units of the neutrino energy scale $\mu_\nu$ since, without reference to external data to fix the energy scale, the prior distribution is invariant against rescaling.  
The predicted distributions are strongly influenced by two boundaries in the parameter space: the first boundary requires the lightest neutrino mass to be positive ($m_L > 0$), which implies 
\begin{align}\label{eq:Sm_min}
	\Sigma m_\nu \ge \Sigma m_\nu^{\rm min} & \equiv \sqrt{ \Delta m_{HM}^2 + \Delta m_{ML}^2} + \sqrt{\Delta m_{ML}^2} \\
	\Sigma m_\nu^{\rm min} & \simeq \begin{cases}
	0.05895 \pm 0.00041 \eV & , \ \text{NO} \\ 
	0.09919 \pm 0.00081 \eV & , \ \text{IO} 
	\end{cases} \nonumber
	\com
\end{align}
where we have used the mass splitting measurements from \sref{sub:Sum_Split}.  
The second boundary is imposed by the step function in \eref{eq:lin_prior}, which implies $\Sigma m_\nu \leq \sqrt{3} \, \mu_\nu$ for the (M) and (D) models.  

Flavor oscillation measurements determine the squared mass splittings, which corresponds to selecting a horizontal section of the parameter space in \fref{fig:2}.  
The (S) model, shown on the right panels of \fref{fig:2}, displays a strong degeneracy:  for any $\mu_\nu$, once the squared mass splittings are fixed, the most probable value of $\Sigma m_\nu$ can be read off from the equality in \eref{eq:Sm_min}, which corresponds to $m_L = 0$.  
For the (D) and (M) models, on the other hand, fixing $\Delta m^2$ leads to a distribution over $\Sigma m_\nu$ that peaks at approximately $\mu_\nu$, regardless of the chosen value for $\Delta m^2$.   

To better understand the distribution over $\Sigma m_\nu$ at fixed splittings, it is useful to consider \fref{fig:3}, which shows the distribution over $\Sigma m_\nu$ for a fixed $\mu_\nu$ with the two squared mass splittings set equal to their best-fit measured values.
The distribution $\pi(\Sigma m_\nu)$ from the semi-analytical analysis is observed to rise linearly and peak at $\Sigma m_\nu = \sqrt{3} \mu_\nu$.  
The linearly rising distribution can be understood from the discussion around \eref{eq:Jnu_SdD} where we argued that $P \propto p(\hat{M}_\nu) \, (\Sigma m_\nu / \mu_\nu)$, which grows linearly with $\Sigma m_\nu$ for flat $p(\hat{M}_\nu)$.  
This argument also implies that NO and IO occur with equal probability if the mass splittings are fixed to their measured values, and $\mu_\nu$ is fixed to a larger value.  
However, the evidence ${\cal E}_{\rm NO} + {\cal E}_{\rm IO}$ drops substantially from $5.6 \times 10^3$ at $\mu_\nu = 0.1 \eV$ to $1.6 \times 10^{-8}$ at $\mu_\nu = 1.0 \eV$ for the (M) model, and it drops from $8.4 \times 10^{1}$ to $2.4 \times 10^{-16}$ for the (D) model.  
The additional suppression for the (D) model is a result of eigenvalue repulsion, and this behavior can be understood analytically.  
The probability density over the summed masses and squared splittings is given by \eref{eq:P_SdD} where the Jacobian determinant $\Jcal_\nu$ is given by \eref{eq:J_nu}; see also \eref{eq:Jnu_SdD}.  
The phenomenon of eigenvalue repulsion is manifest in $\Jcal_\nu$ through the factors of $(m_{\nu i}^2 - m_{\nu j}^2)$.  
The mass dimension of these factors is compensated by factors of $\mu_\nu^{-2}$ through the normalization \pref{eq:Ncal}.  
Taking $\mu_\nu$ to be much larger than the scale of the measured mass splittings results in a suppression that makes such models very improbable; the suppression factors are $\sim (\delta m^2 / \mu_\nu^2)^2 (\Delta m^2 / \mu_\nu^2)^4$ for (D) and $\sim (\delta m^2 / \mu_\nu^2) (\Delta m^2 / \mu_\nu^2)^2$ for (M).  
Thus the overall probability of satisfying the observed mass splittings is extremely small unless $\mu_\nu$ is comparable to the larger splitting.  

\begin{figure*}[t]
\begin{center}
\includegraphics{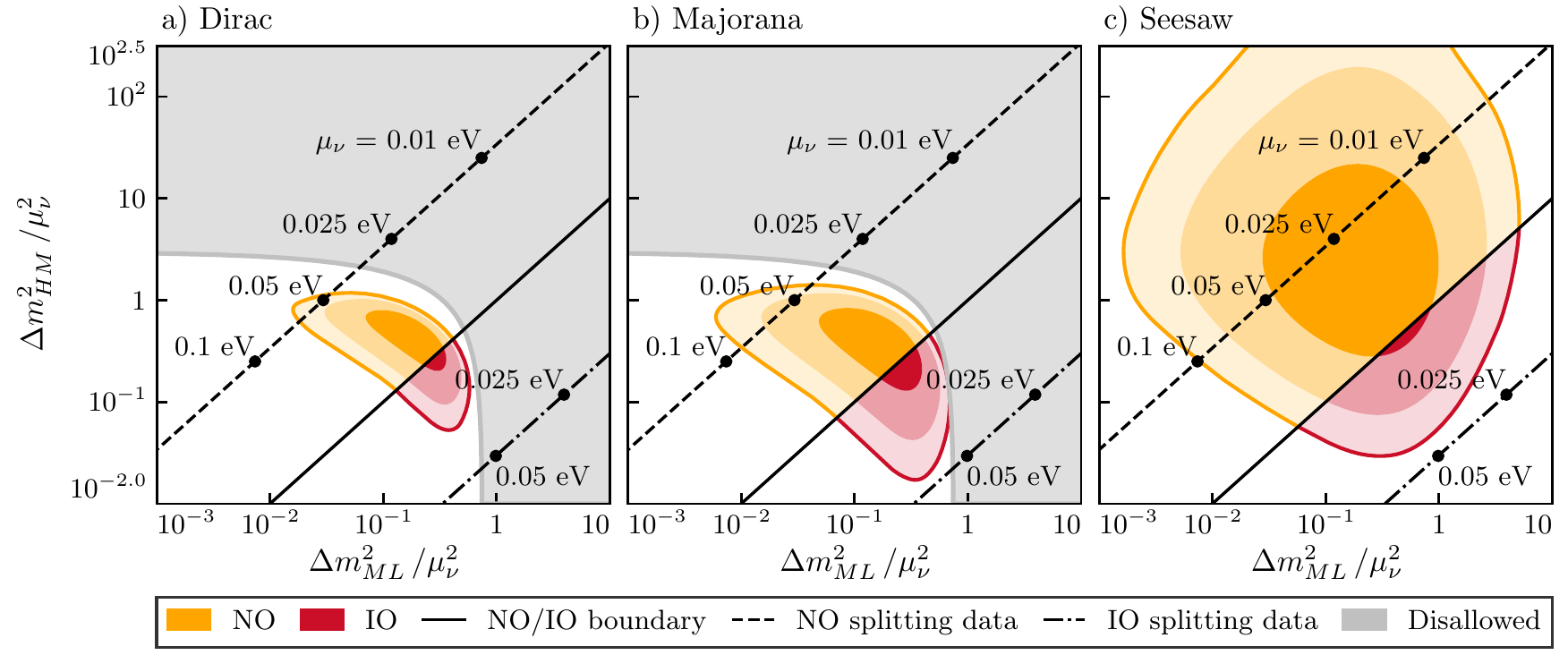} 
\caption{\label{fig:4}
The probability distribution over the squared mass splittings, $\Delta m_{HM}^2$ and $\Delta m_{ML}^2$, after marginalizing over the mass sum, $\Sigma m_\nu$.
The notation and shading are defined in the caption of \fref{fig:2}.  
The solid black line divides models with normal ordering ($\Delta m_{ML}^2 < \Delta m_{HM}^2$) from those with inverted ordering ($\Delta m_{HM}^2 < \Delta m_{ML}^2$).  
Along the black dashed and dot-dashed lines we fix $\Delta m_{HM}^2$ and $\Delta m_{ML}^2$ to equal their measured values (see \sref{sub:Sum_Split}), and we slide the value of $\mu_\nu$.  
}
\end{center}
\end{figure*}

\begin{table*}
\centering
\begin{tabular}{|l|c|c|c|c|c|}
\hline
Model & $\Sigma m_\nu \,[{\rm eV}]$ & $\mu_\nu\,[{\rm eV}]$ & $m_L\,[{\rm eV}]$ & $m_M\,[{\rm eV}]$ & $m_H\,[{\rm eV}]$ \\
\hline
Dirac NO       & $0.069^{+0.01}_{-0.007}$ & $0.055^{+0.006}_{-0.003}$ & $0.008^{+0.007}_{-0.004}$ & $0.01^{+0.005}_{-0.002}$ & $0.0509^{+0.001}_{-0.0008}$ \\
Majorana NO & $0.07^{+0.02}_{-0.01}$ & $0.057^{+0.009}_{-0.006}$ & $0.009^{+0.009}_{-0.006}$ & $0.011^{+0.008}_{-0.002}$ & $0.051^{+0.002}_{-0.001}$ \\
Seesaw NO   & $0.06^{+0.004}_{-0.002}$ & $0.03^{+0.01}_{-0.01}$ & $0.0^{+0.003}_{-0}$ & $0.0087^{+0.0006}_{-0.0005}$ & $0.0505^{+0.0005}_{-0.0005}$ \\
\hline
Dirac IO        & $0.11^{+0.01}_{-0.009}$ & $0.074^{+0.007}_{-0.006}$ & $0.01^{+0.009}_{-0.006}$ & $0.05^{+0.002}_{-0.001}$ & $0.051^{+0.002}_{-0.001}$ \\
Majorana IO  & $0.12^{+0.02}_{-0.01}$ & $0.079^{+0.01}_{-0.009}$ & $0.015^{+0.01}_{-0.01}$ & $0.051^{+0.003}_{-0.002}$ & $0.051^{+0.003}_{-0.002}$ \\
Seesaw IO    & $0.102^{+0.009}_{-0.006}$ & $0.05^{+0.03}_{-0.02}$ & $0.0^{+0.01}_{-0}$ & $0.05^{+0.002}_{-0.002}$ & $0.05^{+0.002}_{-0.002}$ \\
\hline
\end{tabular}
\caption{Marginalized $68\%$ C.L. results on the neutrino prior distribution. Notice that the distributions that are used to extract these bounds are markedly non-Gaussian. Higher confidence regions can be easily computed from the fitting forms in Sec. \ref{sec:Fitting_Formula}.
}
\label{Table:MarginalResults}
\end{table*}

Next we explore the relationship between the ordering, the neutrino energy scale $\mu_\nu$, and the observed splittings.  
In \fref{fig:4} we show the distribution of the two splittings marginalized over $\Sigma m_\nu$ or equivalently over $m_L$, in units of the neutrino mass scale $\mu_\nu$ from the numerical analysis.  
The solid black line divides the regions of parameter space that correspond to the normally ordered mass spectrum (NO) and inverted ordered spectrum (IO).  
In general NO is preferred with respect to IO increasingly from (M) to (D) to (S), as it occupies more volume in parameter space.
In the (M) case this volume factor corresponds to an odd ratio of $4:1$ in favor of NO, in the (D) case of $6:1$ in favor of NO and $22:1$ in the (S) case.  

By studying \fref{fig:4} we can also understand the effect of imposing the mass splitting measurements.  
Fixing the squared mass splittings to equal their measured values selects two solutions in the $(\Delta m_{HM}^2,\Delta m_{ML}^2)$ space, which extend to two lines in the scaled space $(\Delta m_{HM}^2/\mu_\nu^2,\Delta m_{ML}^2/\mu_\nu^2)$ indicated by the dashed line for NO and the dot-dashed line for IO.
Where these lines intersect the high probability regions determines the best values of $\mu_\nu$ for explaining the splittings.  
Now we see that taking $\Sigma m_\nu \gg \sqrt{|\delta m^2|}, \sqrt{|\Delta m^2|}$ selects a model in the low probability tail, which agrees with the behavior seen in \fref{fig:3} that we discussed in the previous paragraph.  
Thus even though such a case would produce a $\Sigma m_\nu$ distribution that peaks at $\sqrt{3} \mu_\nu$ for (D) and (M), it would be a highly unlikely realization of the underlying matrix elements.
Instead the probability is maximal for a value of $\mu_\nu$ that is comparable to scale of the larger mass splitting: $\mu_\nu = 0.02-0.1 \eV$ depending on the scenario.  
For smaller values of $\mu_\nu$ the probability decreases, and it becomes identically zero in the (D) and (M) models because the prior \pref{eq:lin_prior} requires $m_\nu \leq \mu_\nu$.  

\begin{figure*}[t]
\begin{center}
\includegraphics{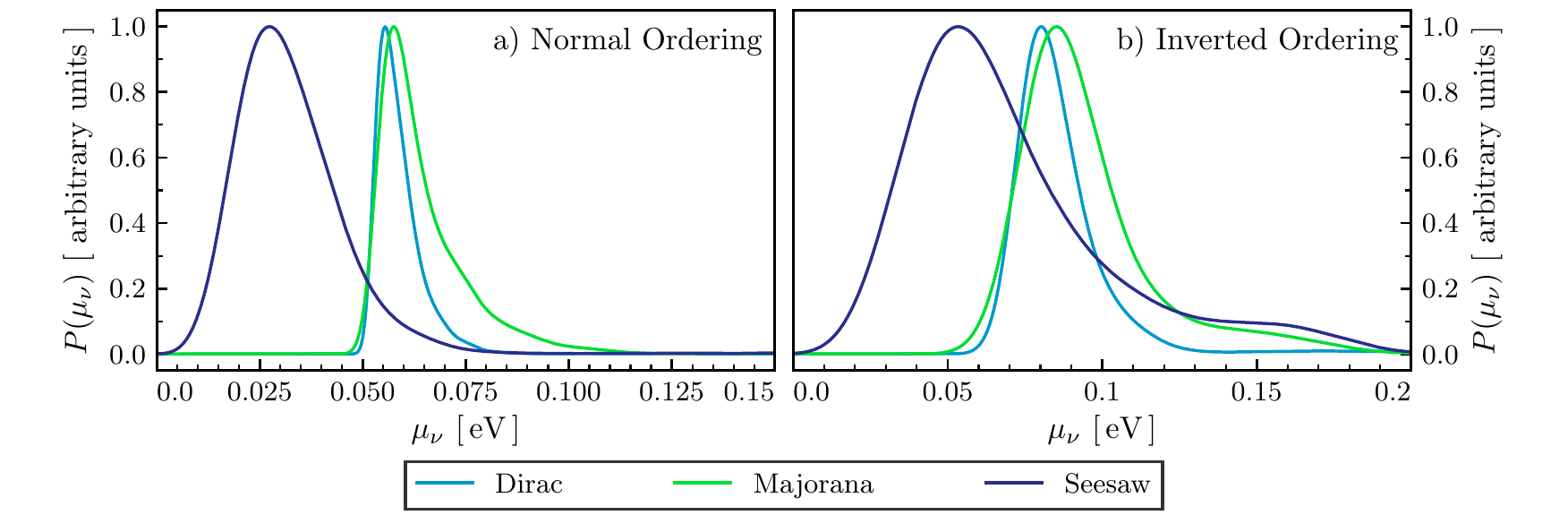} 
\caption{\label{fig:5}
The probability distribution over the neutrino energy scale $\mu_\nu$ after marginalizing over the mass sum and imposing the mass splittings measurements.  
In the left (right) panel we implement the mass splitting measurement with normal (inverted) ordering.
}
\end{center}
\end{figure*}

The key point in the preceding discussion is that in the context of a specific neutrino mass generation mechanism the mass splittings are informative on the neutrino energy scale $\mu_\nu$, because the measured mass splittings set a preferred scale for the mass matrix elements, and thus $\mu_\nu$ should not be chosen {\it a priori}.  
Instead we take it as a hyperparameter and allow the mass splittings to determine its posterior probability distribution.  
We assume a wide, uninformative, flat-linear prior on $\mu_\nu \in (0,1) \eV$ since we have no strong theoretical preference for one value of the neutrino energy scale over another.  
Figure~\ref{fig:5} shows the resulting distribution of $\mu_\nu$ after marginalizing over the other model parameters from the numerical analysis.  
For all three models and both hierarchies, the distribution is sharply peaked around a maximum value that depends on the mechanism and hierarchy.
The confidence levels on the neutrino scale are reported in Table~\ref{Table:MarginalResults}. 

As discussed in \aref{app:Stability} the sharpness of the $\mu_\nu$ peak makes this distribution extremely stable against different prior choices for $\mu_\nu$; for instance, raising the upper limit on $\mu_\nu$ from $1 \eV$ does not affect the resultant distribution on $\Sigma m_\nu$.  
For the Seesaw model one can also specify the hyperprior on the cutoffs of the high-scale matrices, $\mu_D$ and $\mu_M$ from \eref{eq:lin_priorSeesaw}, and let the corresponding distribution on $\mu_\nu$ be inferred from \eref{eq:Seesaw_scale}.  
As long as the resulting distribution on $\mu_\nu$ is uninformative, the distributions over $\Sigma m_\nu$ are unaffected.  
However, the evidences are affected, which complicates model comparison between (S) and either (D) or (M).  

\begin{figure*}[t]
\begin{center}
\includegraphics{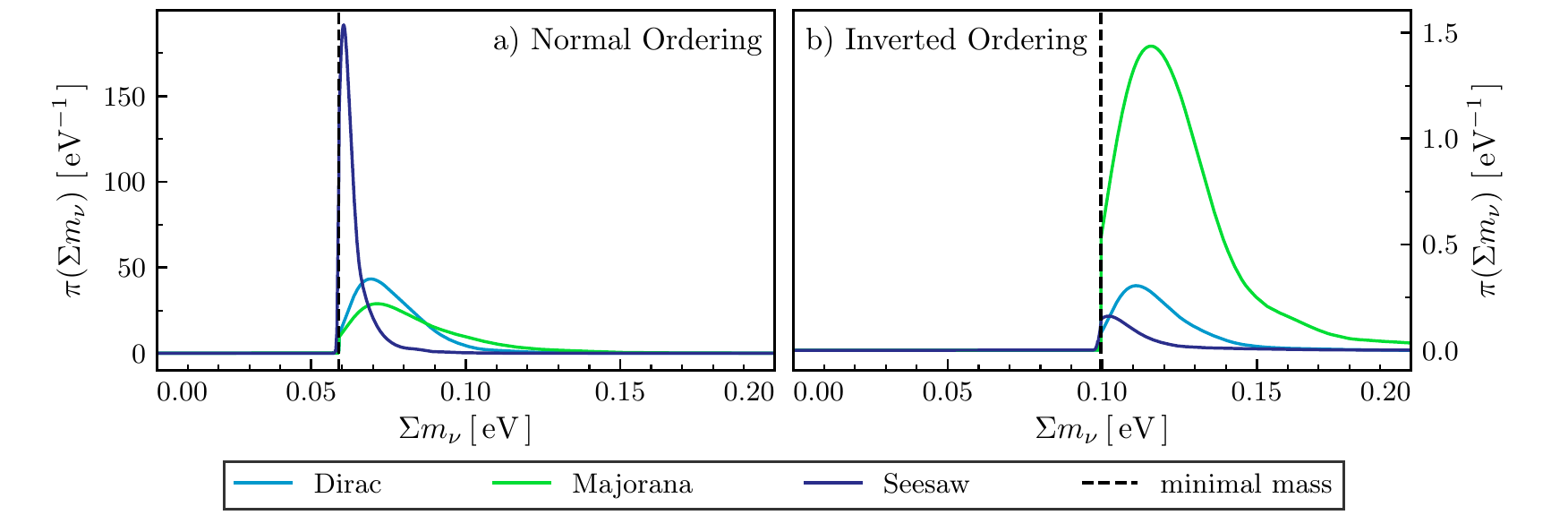} 
\caption{\label{fig:6}
The probability distribution over the neutrino mass sum after marginalizing over the neutrino energy scale $\mu_\nu$ and imposing the squared mass splittings measurements.  
In the left (right) panel we implement the mass splitting measurement with normal (inverted) ordering.  
The vertical black dashed line indicates the minimal value of $\Sigma m_\nu$ consistent with the mass splittings \pref{eq:Sm_min}, which is obtained by setting the mass splittings to their best-fit measured value and setting the lightest neutrino mass to zero. 
}
\end{center}
\end{figure*}

After marginalizing over $\mu_\nu$, the distribution over $\Sigma m_\nu$ from the numerical analysis is shown in \fref{fig:6}.  
This is dramatically different for the (D) and (M) models from the distribution with $\mu_\nu$ held fixed, which we showed in \fref{fig:3}.  
The falling distribution at high $\Sigma m_\nu$ comes from the combination of the $\Sigma m_\nu \le \sqrt{3} \mu_\nu$ limit and the decreasing probability at high $\mu_\nu$ shown in \fref{fig:5}, whereas the peak comes from the fact that for all $\mu_\nu$ consistent with the splittings, the $\Sigma m_\nu$ distribution rises near the minimum.  
In contrast the distribution over $\Sigma m_\nu$ for the (S) model falls off strongly at large $\Sigma m_\nu$ for both fixed $\mu_\nu$ and marginalized $\mu_\nu$.  

The 68\% confidence levels on the neutrino mass sum and other parameters from the numerical analysis are reported in \tref{Table:MarginalResults}.  
Evidences are reported in \tref{Table:FittingFormula} and come from the semi-analytical calculation for (M) and (D) and the numerical multinest calculation for (S). 

We notice that once splitting measurements are factored into our study, the preference for NO with respect to IO becomes higher; the evidence increases by $\Delta \log_{10} \mathcal{E} = 2.1$ for (D), $\Delta \log_{10} \mathcal{E} = 1.3$ for (M), and $\Delta \log_{10} \mathcal{E} = 2.7$ for (S). 
These odds are easily understood from Figs.~\ref{fig:4}~and~\ref{fig:5}.
Consider first the (D) model with the splittings fixed to their measured values and NO assumed (dashed line in \fref{fig:4}).  
Figure~\ref{fig:5} shows that the most probable value for the neutrino energy scale is $\mu_\nu \approx 0.06 \eV$, which lies between the $2\sigma$ ($95\%$ C.L.) and $3\sigma$ ($99.7\%$ C.L.) contours in \fref{fig:4}.  
On the other hand, if IO is assumed (dot-dashed line), the most probable value is $\mu_\nu \approx 0.08 \eV$ from the right panel of \fref{fig:5}, and it lies outside of the $3\sigma$ contour and closer to the $4\sigma$ ($0.99993\%$ C.L.) one in \fref{fig:4}.  
The probability of these two events explains the approximately $130:1$ odds favoring NO over IO that was reported above.  
Consider next the (M) model for which the most probable neutrino energy scales are $\mu_\nu \approx 0.06 \eV$ for NO and $\mu_\nu \approx 0.08 \eV$ for IO, which correspond to points in \fref{fig:4} that are at the $2\sigma$ and $3\sigma$ levels, respectively.
The relative probability of these two events explains the $19:1$ odds in favor of NO for (M).  
In (S) the maximal $\mu_\nu$ occurs at $\mu_\nu \approx 0.025 \eV$ for NO and at $\mu_\nu \approx 0.05 \eV$ for IO, and the corresponding points in \fref{fig:4} are at approximately the $0.5\sigma$ level (NO) and slightly above $3\sigma$ (IO).
This explains the odds of approximately $470:1$ in favor of NO for (S). 

The evidence results in \tref{Table:MarginalResults} also allows us to comment on the odds of different models summed over the two hierarchies.  
Given equal prior odds of S:M:D, conditioning to the mass splitting data gives us odds of $830:3.3:1$.  

\subsection{Fitting formulas}\label{sec:Fitting_Formula}

In this section we present an empirical fitting formula that approximates the probability distribution over $\Sigma m_\nu$ for practical applications.
This is constructed from three building blocks. 

The first component is a smooth cutoff at the minimal value of the sum of the masses given by \eref{eq:Sm_min}:  $x_{\rm min} = 0.05895 \eV$  ($x_{\rm min} = 0.09919 \eV$) with width $\sigma_{\rm min} = 0.00041 \eV$ ($\sigma_{\rm min} = 0.00081 \eV$) for NO (IO).  
The cutoff is implemented as a smoothed step function: 
\begin{align}
 {\rm Step}(x,\mu, \sigma) \equiv \frac{1}{2}\left[ 1+ {\rm Erf}\left( \frac{x-\mu}{\sqrt{2}\sigma}\right) \right] \com
\end{align}
where $\mu$ is the value at which the function transitions from zero to one, and $\sigma$ is the width of the transition. 

The second component is a skewed Gaussian distribution that describes the behavior of the peak of the distribution.  
The center of the Gaussian is denoted by $\bar{x}$, its width is $\sigma_{\bar{x}}$, and the skewness is $\alpha$. 

The third component is a power law cutoff at large $\Sigma m_\nu$.  
To prevent this component from dominating at small mass, it is cut off using a smooth transition centered at $\hat{x}$ with width $\sigma_{\hat{x}}$.
The exponent of the power law cutoff is taken to be $p=-16$, $-10$, and $-12$ for the Dirac, Majorana, and Seesaw models, respectively.  

Joining these pieces together we obtain a fitting function for the prior $\pi(\Sigma m_\nu)$:  
\begin{align}\label{eq:pi_fit}
	P_{\rm fit}(x) & = N \cdot \, {\rm Step}(x,x_{\rm min}, \sigma_{\rm min} ) \nonumber \\
& \quad  \bigg[  \frac{\sqrt{2}}{\sqrt{\pi\sigma_{\bar{x}}^2}}\,{\rm Step}(x,\bar{x}, \alpha ) \, {\rm exp} \left( -\frac{(x-\bar{x})^2}{2\sigma_{\bar{x}}^2} \right)
\nonumber \\
& \qquad + \mathcal{A} \, {\rm Step}(x,\hat{x}, \sigma_{\hat{x}} )\, \left( \frac{x}{\hat{x}}\right)^{p}\bigg] \per
\end{align}
The relative amplitude between the exponential and power law behavior is given by $\mathcal{A}$.
The coefficient $N$ ensures that the fitting function is normalized such that $\int_{0}^{\infty} \ud x \, P_{\rm fit}(x) = 1$.  

\begin{figure*}[t]
\begin{center}
\includegraphics{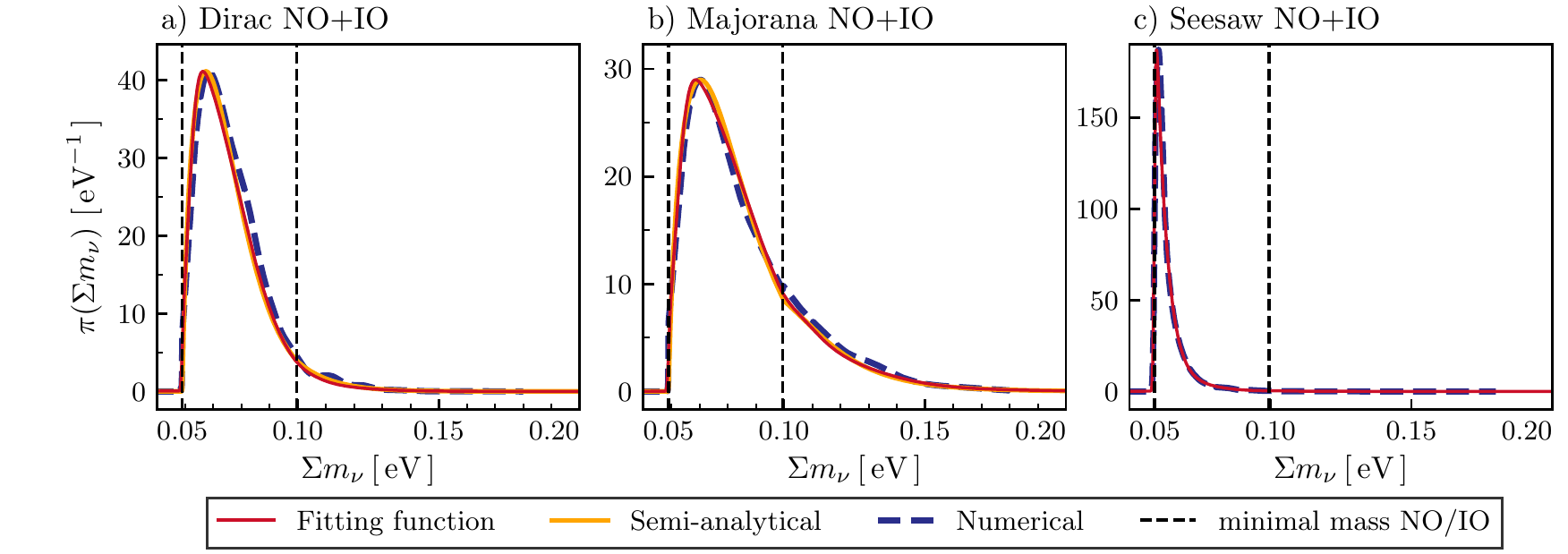} 
\caption{\label{fig:7}
The probability distribution over the neutrino mass sum after marginalizing over the neutrino energy scale $\mu_\nu$, marginalizing over the ordering of the mass spectrum, and imposing the mass splitting measurements.  
For the Dirac and Majorana models, there is an excellent agreement between the semi-analytical calculation and the fully numerical MCMC results; a semi-analytic calculation was not performed for the Seesaw model.  
In all three cases, the empirical fitting function also matches extremely well.  
}
\end{center}
\end{figure*}

Table \ref{Table:FittingFormula} summarizes the parameters of the fitting formulas for the three different models considered in \sref{sec:Lin_Prior} (Dirac, Majorana, Seesaw) and the two spectrum orderings (NO, IO).  
For the (D) and (M) models we have fit to the semi-analytical results, since they are more accurate, and for the (S) model we fit to the numerical results.  
A reference implementation of the fitting formulas can be found at \url{https://github.com/mraveri/Neutrino_Prior}.
For all six cases we observe a very good agreement between the semi-analytical results, the numerical results, and the fitting formula; this agreement is illustrated in \fref{fig:7}.  
In the (M) and (D) cases, the fitting formula agrees with the semi-analytical results at $1\%$ on the $95\%$ C.L. intervals and at $5\%$ on the $0.99993$ C.L. ($4\sigma$) intervals; in the (S) case, the fitting formula agrees with the numerical results at $1\%$ on the $95\%$ C.L. intervals and at $10\%$ on the $0.99993$ C.L. ($4\sigma$) tails, but note that the numerical results themselves become increasingly poorly sampled in the tails. 

Table~\ref{Table:FittingFormula} summarizes the parameters of the fitting formulas for six different combinations of orderings and mass models. 
The reader may choose to combine these fitting functions by marginalizing over orderings or models.  
The (totally) marginalized probability distribution is calculated as 
\begin{equation}\label{eq:pi_fit_tot}
	P_{\rm fit, tot}(x) = \frac{\sum_i {\cal E}_i \, \Pi_i \, P_{\rm fit}(x | i )}{\sum_i {\cal E}_i \, \Pi_i} \,,
\end{equation}
where $P_{\rm fit}(x | i )$ is given by \eref{eq:pi_fit} for one of the rows of \tref{Table:FittingFormula}, the evidences $\Ecal_i$ appear in \tref{Table:FittingFormula}, and we allowed for the possibility of having a prior $\Pi_i$ on the $i$-th model.  
Note that within each mass model (Dirac, Majorana, or Seesaw), we have assumed $\Pi_{\rm IO}/\Pi_{\rm NO}=1$.  
Then to obtain the prior $\pi(\Sigma m_\nu)$ appearing in \fref{fig:6}, it is necessary to scale the fitting function by $\Ecal_{\rm NO} / \bigl( \Ecal_{\rm NO} + \Ecal_{\rm IO} \bigr)$ for the NO distribution and by $\Ecal_{\rm IO} / \bigl( \Ecal_{\rm NO} + \Ecal_{\rm IO} \bigr)$ for the IO distribution.  

\begin{table*}
\centering
\begin{tabular}{|l|c|c|c|c|c|c|c|c|c|}
\hline
Model &  ${\rm log}_{10} \mathcal{E}$ & ${\rm log}_{10} \,N $ & $\bar{x}~[{\rm eV}]$ & $\sigma_{\bar{x}}~[{\rm eV}]$ & $\alpha~[{\rm eV}]$ & ${\rm log}_{10} \,(\mathcal{A}~{\rm eV})$ & $\hat{x}~[{\rm eV}]$ & $\sigma_{\hat{x}}~[{\rm eV}]$ & $p$ \\
\hline
Dirac NO       & $3.08$ & $-0.006035$ & $0.06109$ & $0.01700$ & $0.002484$ & $-0.6430$ & $0.1315$ & $0.01603$ &  $-16$ \\
Majorana NO & $3.58$ & $-0.01028$ & $0.06164$ & $0.02416$ & $0.003215$ & $0.1945$ & $0.1336$ & $0.01251$ & $-10$ \\
Seesaw NO   & $6.00$ & $26.91$ & $-0.4099$ & $0.04301$ & $0.01624$ & $-26.25$ & $0.0779$ & $0.007016$ & $-12$ \\
\hline
Dirac IO        & $0.96$ & $-0.01345$ & $0.1033$ & $0.01792$ & $0.002936$ & $-0.2773$ & $0.1713$ & $0.01876$ & $-16$ \\
Majorana IO  & $2.31$ & $-0.02738$ & $0.1039$ & $0.02490$ & $0.003634$ & $0.4442$ & $0.1705$ & $0.01359$ & $-10$ \\
Seesaw IO    & $3.33$ &  $65.07$ & $0.0000$ & $0.01851$ & $-0.006123$ & $-62.94$ & $0.09837$ & $0.003094$ & $-12$ \\
\hline
\end{tabular}
\caption{Parameters for the fitting formula discussed in Sec. \ref{sec:Fitting_Formula}. 
We also report in the first column the model prior volume, without considering splitting data, and, in the second column, the overall normalization of the prior distributions once splitting data are included and we marginalize over $\mu_\nu$. 
}
\label{Table:FittingFormula}
\end{table*}

In \fref{fig:7} we show the distribution of $\Sigma m_\nu$ after marginalizing over the two hierarchy to form the total distribution in the (D,M,S) models.  
We also  showcase here the agreement between different ways of obtaining such distributions by plotting the fitting formula against the semi-analytical results and the numerical ones.
This total prior can then be used in cosmological parameter estimation.  

As an example let us return to the test case considered in \sref{sec:AdHoc}.  
Recall that this test case involved cosmological data that constrained $\Sigma m_\nu$ to an uncertainty of $\sigma =0.3 \eV$ around zero.  
If we repeat the same calculation for the priors in \tref{sec:Fitting_Formula} instead, we obtain the following $95\%$ CL upper limits: $\Sigma m_\nu < 0.0975 \eV$ for (D), $0.126 \eV$ for (M), and $0.0730 \eV$ for (S) after marginalizing over the ordering with \eref{eq:pi_fit_tot}.  
Since the distributions are sharply peaked toward small masses, the inferred $95\%$ CL upper limits are stronger than the ones we obtained for the flat-linear and flat-log priors in \sref{sec:AdHoc}.  \nl

\section{Conclusion}\label{sec:Conc}

In this work we have discussed the choice of priors on the sum of neutrino masses $\Sigma m_\nu$ for cosmological data analyses.  
Whereas it is customary to assume a flat prior on $\Sigma m_\nu$, which is an ad hoc choice, we have instead argued that the physically motivated choice is to specify the prior at the level of the neutrino mass matrix $M_\nu$.  
In this regard, our study extends earlier work on neutrino flavor model building with the anarchy hypothesis and applies that formalism to cosmological observables.  
Specifically, we focus on the basis-independent anarchy hypothesis (BAH) that assigns equal probability to any matrices that can be related by a change of basis.  
Subject to the BAH restriction, simple priors on $M_\nu$ include a flat or Gaussian distribution on its individual matrix elements.  

One of the main conclusions of our work, is that these simple implementations of the BAH generally disfavor the degenerate regime where the neutrino mass scale is much larger than the mass splittings, $\Sigma m_\nu \gg \sqrt{|\delta m^2|}, \sqrt{|\Delta m^2|}$.  
This is because the probability distribution over elements of the neutrino mass matrix $M_\nu$ necessarily selects a scale $\mu_\nu$, e.g. the cutoff on a flat distribution or the variance of a Gaussian distribution.  
Due to the phenomenon of eigenvalue repulsion, high probability spectra have $\Sigma m_\nu \sim \sqrt{|\delta m^2|} \sim \sqrt{|\Delta m^2|} \sim \mu_\nu$, and in particular, degenerate spectra are especially unlikely.  
The repulsion effect can be understood from the Jacobian determinant that relates the distribution over (mass) matrix elements to the distribution over (mass) eigenvalues.  
This determinant [see \eref{eq:J_nu} or \pref{eq:Jnu_SdD}] is proportional to $|\delta m^2|$ and $|\Delta m^2|$, which makes the probability density proportional to $|\delta m^2| / \mu_\nu^2$ and $|\Delta m^2| / \mu_\nu^2$.  
Then in order for the measured mass splittings to be a probable realization, it is necessary to take $\mu_\nu \sim \sqrt{|\Delta m^2|} \sim 0.05 \eV$, see \fref{fig:5}, which also implies $\Sigma m_\nu \sim 0.05 \eV$, see \fref{fig:6}.  
Consequently, simple basis-independent priors at the level of the neutrino mass matrix translate into distributions over $\Sigma m_\nu$ that peak around the smallest value allowed by the measured mass splittings \pref{eq:Sm_min}, roughly $0.06 \eV$ for normal ordering and $0.10 \eV$ for inverted ordering.  

Using the same reasoning as above, one can see that a flat distribution over $\Sigma m_\nu$, which is often assumed in cosmological studies, is highly improbable for {\it simple} basis-independent priors on $M_\nu$.  
In other words, this ad hoc prior assumption is not well motivated from fundamental physical principles described by random, anarchical neutrino mass matrices.  

In order to obtain qualitatively different results, it is necessary to chose a prior on $M_\nu$ that is not {\it simple}.  
As we discuss in \sref{sec:AdHocRevisited}, one can counterbalance the effects of eigenvalue repulsion by choosing the prior on the neutrino mass matrix to be the reciprocal of the Jacobian determinant.  
Although this leads to a wider tail in the $\Sigma m_\nu$ distribution (see  \aref{app:Stability}), now the prior is expressed as a very complicated function of the neutrino mass matrix, which undercuts the physical motivation for the anarchy hypothesis.  

Our main quantitative results appear in \sref{sec:Priors} where we focus on a particular prior distribution that is flat in the individual elements of the neutrino mass matrix out to a scale $\mu_\nu$ for the Dirac and Majorana models, and it is flat in the separate high-scale Dirac and Majorana matrices for the Seesaw model.  
We derive the distributions of the sum of neutrino masses and the squared mass splittings, and our results are discussed in \sref{sec:Lin_Prior}.  
Most notably, the predicted distribution over the sum of neutrino masses $\Sigma m_\nu$ appears in \fref{fig:6}.  
For all three models (Dirac, Majorana, and Seesaw) the distribution is sharply peaked close to the lowest value allowed by the measured neutrino mass splittings, but the models are notably distinguished by the behavior in the tail of the distribution at high $\Sigma m_\nu$.  
We find that this prior prefers the mass spectrum with normal ordering over inverted ordering with odds $130 : 1$ for (D), $19 : 1$ for (M), and $470 : 1$ for (S); see also \fref{fig:4} [cf., \rref{Simpson:2017qvj}].  
We thoroughly tested the stability of these results under the various assumptions that are used to build the prior distributions; for a detailed discussion see \aref{app:Stability}.  

The reader is encouraged to apply the prior probabilities $\pi(\Sigma m_\nu)$ appearing in \fref{fig:6} to cosmological data analyses of other cosmological parameters, and an empirical fitting formula is available in \sref{sec:Fitting_Formula}.  
Since the distributions in \fref{fig:6} are sharply peaked, this offers some justification for cosmological studies that simply fix $\Sigma m_\nu$ to equal the minimal value consistent with the measured squared mass splittings \pref{eq:Sm_min}, roughly $0.06 \eV$ for NO and $0.1 \eV$ for IO.  
In fact the priors that we have presented here may be useful from a phenomenological perspective if one seeks to have a prior that favors minimal mass but also allows for the possibility that a true preference by the data may drive the fit to larger $\Sigma m_\nu$.  
For practical applications, we recommend the reader start with the prior for the Majorana model, since it is least sharply peaked.  

Let us close by discussing the impact of our analysis for cosmological probes of neutrino mass and their potential implications for neutrino mass models. 
Given the current sensitivity of the cosmological measurements, the priors $\pi(\Sigma m_\nu)$ discussed here are presently more informative than the data.  
In this sense the priors can be viewed as providing targets for future experimental searches.
In particular, our priors define several challenging objectives of increasing experimental sensitivity, which can be inferred from marginalized distributions in Figs.~\ref{fig:6}~and~\ref{fig:7} and their corresponding fitting functions in \sref{sec:Fitting_Formula}.  
If you detect $\Sigma m_\nu \gtrsim 0.13 \eV$ with sufficient experimental accuracy\footnote{Here we mean that the experimental uncertainty is small in the sense that $\sigma(\Sigma m_\nu) \ll \bigl( \Sigma m_\nu^{99.7} - \Sigma m_\nu^{\rm min} \bigr) / 3$ where $\Sigma m_\nu^{\rm min}$ is given by \eref{eq:Sm_min} and  $\Sigma m_\nu^{99.7}$ is the quoted 99.7\% CL threshold that gives the 3$\sigma$ Gaussian equivalent width of the prior distribution from the minimum.  Otherwise, the experimental error should be combined appropriately
into a joint significance.} then you rule out the Dirac model (i.e., it is disfavored at greater than $99.7\%$ confidence); if you detect $\Sigma m_\nu \gtrsim 0.18 \eV$ then you rule out the Majorana model; and if you detect $\Sigma m_\nu \gtrsim 0.10 \eV$ then you rule out the Seesaw model.  
By the time when the cosmological measurements reach this level of sensitivity to $\Sigma m_\nu$, the ordering of the neutrino mass spectrum may already be known from terrestrial experiments.  
If the neutrino mass spectrum has normal (inverted) ordering, and you detect $\Sigma m_\nu \gtrsim 0.12 \eV$ ($0.18 \eV$) then you rule out the Dirac model; if you detect $\Sigma m_\nu \gtrsim 0.17 \eV$ ($0.23 \eV$) then you rule out the Majorana model; and if you detect $\Sigma m_\nu \gtrsim 0.097 \eV$ ($0.17 \eV$) then you rule out the Seesaw model.  
Recall however that the odds for normal versus inverted ordering in each case already disfavors any mass value for the latter case.  

In this way, increasing accuracy to $\Sigma m_\nu$ from cosmological observations will test the hypothesis that neutrino masses arise from unspecified high energy physics that can be described by effectively random mass matrices distributed such that matrices related by a change of flavor basis are equally probable.  
As we have explored in this work, the phenomenon of eigenvalue repulsion in the simplest implementation of the basis-independent anarchy hypothesis requires $\Sigma m_\nu$ to be not much larger than the scale of the larger squared mass splittings, $\sqrt{\Delta m^2} \sim 0.05 \eV$, which leads to a tension with the data if $\Sigma m_\nu$ is measured to be larger (see previous paragraph).  
If next-generation experiments measure a large $\Sigma m_\nu$, what does this imply for the basis-independent anarchy hypothesis?  
In such a situation, one is forced to abandon the simplest implementation of the BAH  that we consider in this work.
One can nevertheless construct potentially viable models if the prior on the mass matrix is chosen to counterbalance the eigenvalue repulsion, but such baroque priors are not in the {\it spirit} of the anarchy hypothesis and they undercut its physical motivation.  

\begin{acknowledgements}
We are grateful to Jonathan Braden, Stephen Feeney, David J. E. Marsh, Eugene Lim, Hitoshi Murayama, Serguey Petcov, and Lian-Tao Wang for discussions.  
AJL and WH are supported at the University of Chicago by the Kavli Institute for Cosmological Physics through grant NSF PHY-1125897 and an endowment from the Kavli Foundation and its founder Fred Kavli.  
MR and WH are supported by U.S. Dept. of Energy contract DE-FG02-13ER41958.
WH was additionally supported by NASA ATP NNX15AK22G and the Simons Foundation.  
\end{acknowledgements}

\vspace{-0.2cm} 
\appendix

\section{Stability of the prior on $\Sigma m_\nu$}\label{app:Stability}
In order to assess the stability of our main results, which appear in \sref{sec:Lin_Prior} and specifically \fref{fig:6}, we enumerate here the various assumptions, and we test the effect of relaxing or modifying each assumption.  
Overall we find that the prior $\pi(\Sigma m_\nu)$ is very robust to what we called {\it simple} priors in the main text.  

In calculating $\pi(\Sigma m_\nu)$, which appears in \fref{fig:6}, we have made two key assumptions: the probability distribution is assumed to be flat in the matrix elements with a cutoff at ${\rm Tr} \, M_\nu^\dagger M_\nu = \mu_\nu^2$ \pref{eq:lin_prior}; the cutoff $\mu_\nu$ is marginalized with a flat prior from $0 \eV$ to $1 \eV$.   

\begin{figure}[t]
\begin{center}
\includegraphics{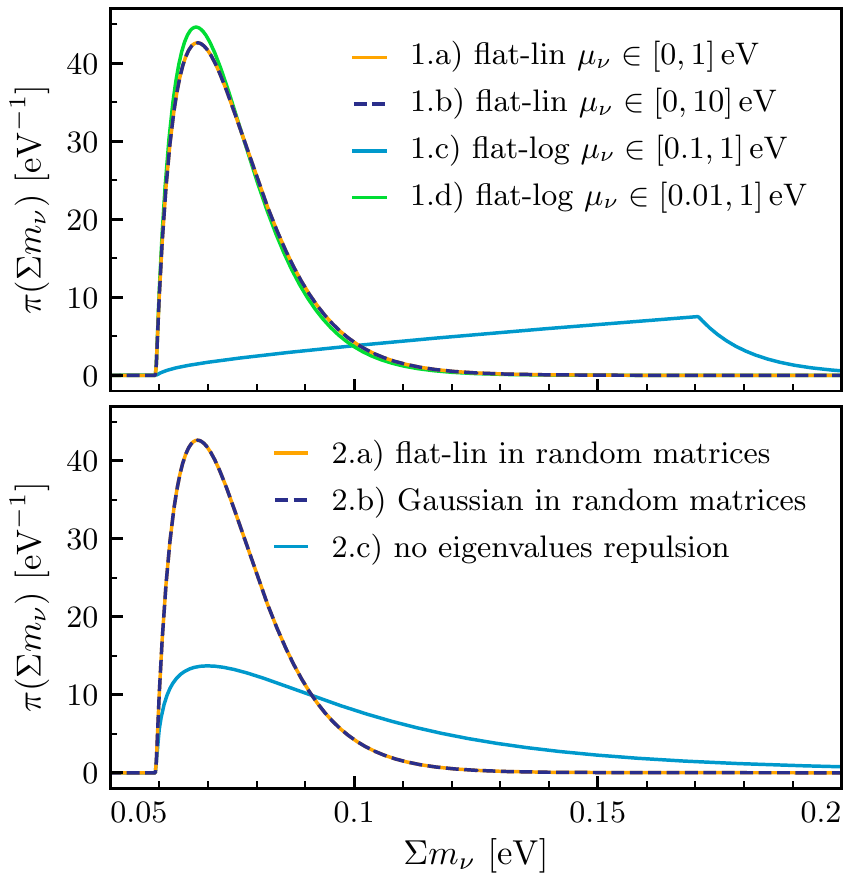} 
\caption{\label{fig:8}
Stability of the prior $\pi(\Sigma m_\nu)$ against the various assumptions.
}
\end{center}
\end{figure}

Let us first discuss how our results depend upon the prior on the neutrino energy scale $\mu_\nu$.  
As we saw in \fref{fig:5}, the distribution over $\mu_\nu$ is sharply peaked.  
As such it is difficult to change the distribution $\pi(\Sigma m_\nu)$ with relatively uninformative priors on $\mu_\nu$.  
To demonstrate this point we show in \fref{fig:8} the distribution $\pi(\Sigma m_\nu)$ calculated from various different priors on $\mu_\nu$, namely (1.a) a flat prior on $\mu_\nu \in (0 \eV, 1 \eV)$, (1.b) a flat prior on $\mu_\nu \in (0 \eV, 10 \eV)$, (1.c) a flat prior on $\log \mu_\nu \in (\log 0.1 \eV, \log 1 \eV)$, and (1.d) a flat prior on $\log \mu_\nu \in (\log 0.01 \eV, \log 10 \eV)$.
In these cases the evidences for the models read: $\log_{10} \mathcal{E}(a)=3.1$, $\log_{10} \mathcal{E}(b)=2.1$, $\log_{10} \mathcal{E}(c)=0.14$, and $\log_{10} \mathcal{E}(d)=3.5$.  
As we can see in \fref{fig:8} (1.a,1.b,1.d), once the $\mu_\nu$ prior encompasses the region where the $\mu_\nu$ posterior peaks, without adding significant curvature, the prior distributions $\pi(\Sigma m_\nu)$ are almost identical.  
Case (1.a) is a factor ten preferred to (1.b) as the latter has wider prior on $\mu_\nu$ that do not contain significant posterior.  
Case (1.d) is a factor two preferred with respect to case (1.a) as it assigns more weight to the region of smaller $\mu_\nu$ where the posterior is peaked.
When the prior on $\mu_\nu$ excludes the region of maximum posterior, as in \fref{fig:8} (1.c), and selects out a particular value of $\mu_\nu$, the shape of the distribution is changed but the model has overall smaller probability with odds $800:1$ with respect to case (1.a). 
In this case the hyperprior on $\mu_\nu$ counteracts the natural tendency from eigenvalue repulsion that favors $\Sigma m_\nu \sim  \sqrt{|\Delta m^2|} \sim \mu_\nu \sim 0.05 \eV$ but also makes the observed splittings relatively unlikely; see \sref{sec:Conc}.  
While the figure only shows the (D) model, similar results hold for the (M) and the (S) models, relying on the peaked structure of the $\mu_\nu$ posterior.
In other words, the prior distribution $\pi(\Sigma m_\nu)$ is insensitive to the prior on $\mu_\nu$ as long as these priors on $\mu_\nu$ have negligible curvature around the region of parameter space where the distribution of $\mu_\nu$ peaks.
Notice that in the (S) model a flat-log prior on the effective neutrino energy scale corresponds to a log prior on both the seesaw energy scales [$\mu_D$ and $\mu_M$ in \eref{eq:lin_priorSeesaw}]. 

Let us next discuss how our results depend upon the prior over neutrino mass matrix elements.
We also show in \fref{fig:8} the distribution $\pi(\Sigma m_\nu)$ calculated from various different priors on $M_\nu$, namely (2.a) the flat prior $p_\nu(\hat{M}_\nu \, | \, \mu_\nu) \propto \Theta\bigl( \mu_\nu^2 - {\rm Tr}\bigl[ \hat{M}_\nu^\dagger \hat{M}_\nu \bigr] \bigr)$ from \eref{eq:lin_prior}, (2.b) a Gaussian prior $p_\nu(\hat{M}_\nu \, | \, \mu_\nu) \propto \, {\rm exp}\bigl[ - {\rm Tr}[M_\nu^\dagger M_\nu] / 2 \mu_\nu^2 \bigr]$, and (2.c) a prior that balances eigenvalue repulsion $p_\nu(\hat{M}_\nu \, | \, \mu_\nu) \propto \, | \Jcal_\nu(\hat{M}_\nu) |^{-1} ( \det M_\nu )^{2/3}$ that was discussed in \sref{sec:AdHocRevisited}.  
We marginalize each distribution over the neutrino energy scale with a flat distribution $\mu_\nu \in (0, 1) \eV$.  
For these cases the evidences are $\log_{10} \mathcal{E}(a)=3.1$, $\log_{10} \mathcal{E}(b)=2.4$, $\log_{10} \mathcal{E}(c)=4.1$.  
The distribution $\pi(\Sigma m_\nu)$ is nearly identical for the flat and Gaussian priors, although the evidence prefers the flat case.
On the other hand, the third prior balances the eigenvalue repulsion factors in $\Jcal_\nu \sim | m_{\nu i}^2 - m_{\nu j}^2 |$ by taking $p_\nu \propto |\Jcal_\nu|^{-1}$, which increases the evidence, and the additional factor of $p_\nu \propto (\det M_\nu)^{2/3}$ enhanced the high mass tail.  
In summary, the results presented already in \sref{sec:Lin_Prior} are robust against changing the prior on $M_\nu$ due to the presence of eigenvalue repulsion, which prefers the mass scale of the prior to lie close to the mass scale of the measured neutrino mass splittings.  
In order to obtain qualitatively different results, it is necessary to choose a prior on $M_\nu$ that counterbalances the eigenvalue repulsion.   

\bibliography{refs--NuPriors}

\begin{thebibliography}{47}%
\makeatletter
\providecommand \@ifxundefined [1]{%
 \@ifx{#1\undefined}
}%
\providecommand \@ifnum [1]{%
 \ifnum #1\expandafter \@firstoftwo
 \else \expandafter \@secondoftwo
 \fi
}%
\providecommand \@ifx [1]{%
 \ifx #1\expandafter \@firstoftwo
 \else \expandafter \@secondoftwo
 \fi
}%
\providecommand \natexlab [1]{#1}%
\providecommand \enquote  [1]{``#1''}%
\providecommand \bibnamefont  [1]{#1}%
\providecommand \bibfnamefont [1]{#1}%
\providecommand \citenamefont [1]{#1}%
\providecommand \href@noop [0]{\@secondoftwo}%
\providecommand \href [0]{\begingroup \@sanitize@url \@href}%
\providecommand \@href[1]{\@@startlink{#1}\@@href}%
\providecommand \@@href[1]{\endgroup#1\@@endlink}%
\providecommand \@sanitize@url [0]{\catcode `\\12\catcode `\$12\catcode
  `\&12\catcode `\#12\catcode `\^12\catcode `\_12\catcode `\%12\relax}%
\providecommand \@@startlink[1]{}%
\providecommand \@@endlink[0]{}%
\providecommand \url  [0]{\begingroup\@sanitize@url \@url }%
\providecommand \@url [1]{\endgroup\@href {#1}{\urlprefix }}%
\providecommand \urlprefix  [0]{URL }%
\providecommand \Eprint [0]{\href }%
\providecommand \doibase [0]{http://dx.doi.org/}%
\providecommand \selectlanguage [0]{\@gobble}%
\providecommand \bibinfo  [0]{\@secondoftwo}%
\providecommand \bibfield  [0]{\@secondoftwo}%
\providecommand \translation [1]{[#1]}%
\providecommand \BibitemOpen [0]{}%
\providecommand \bibitemStop [0]{}%
\providecommand \bibitemNoStop [0]{.\EOS\space}%
\providecommand \EOS [0]{\spacefactor3000\relax}%
\providecommand \BibitemShut  [1]{\csname bibitem#1\endcsname}%
\let\auto@bib@innerbib\@empty
\bibitem [{\citenamefont {Ade}\ \emph {et~al.}(2014)\citenamefont {Ade} \emph
  {et~al.}}]{Ade:2013zuv}%
  \BibitemOpen
  \bibfield  {author} {\bibinfo {author} {\bibfnamefont {P.~A.~R.}\
  \bibnamefont {Ade}} \emph {et~al.} (\bibinfo {collaboration} {Planck}),\
  }\href {\doibase 10.1051/0004-6361/201321591} {\bibfield  {journal} {\bibinfo
   {journal} {Astron. Astrophys.}\ }\textbf {\bibinfo {volume} {571}},\
  \bibinfo {pages} {A16} (\bibinfo {year} {2014})},\ \Eprint
  {http://arxiv.org/abs/1303.5076} {arXiv:1303.5076 [astro-ph.CO]} \BibitemShut
  {NoStop}%
\bibitem [{\citenamefont {Ade}\ \emph {et~al.}(2016)\citenamefont {Ade} \emph
  {et~al.}}]{Ade:2015xua}%
  \BibitemOpen
  \bibfield  {author} {\bibinfo {author} {\bibfnamefont {P.~A.~R.}\
  \bibnamefont {Ade}} \emph {et~al.} (\bibinfo {collaboration} {Planck}),\
  }\href {\doibase 10.1051/0004-6361/201525830} {\bibfield  {journal} {\bibinfo
   {journal} {Astron. Astrophys.}\ }\textbf {\bibinfo {volume} {594}},\
  \bibinfo {pages} {A13} (\bibinfo {year} {2016})},\ \Eprint
  {http://arxiv.org/abs/1502.01589} {arXiv:1502.01589 [astro-ph.CO]}
  \BibitemShut {NoStop}%
\bibitem [{\citenamefont {Riemer-Sorensen}\ \emph {et~al.}(2012)\citenamefont
  {Riemer-Sorensen} \emph {et~al.}}]{RiemerSorensen:2011fe}%
  \BibitemOpen
  \bibfield  {author} {\bibinfo {author} {\bibfnamefont {S.}~\bibnamefont
  {Riemer-Sorensen}} \emph {et~al.},\ }\href {\doibase
  10.1103/PhysRevD.85.081101} {\bibfield  {journal} {\bibinfo  {journal} {Phys.
  Rev.}\ }\textbf {\bibinfo {volume} {D85}},\ \bibinfo {pages} {081101}
  (\bibinfo {year} {2012})},\ \Eprint {http://arxiv.org/abs/1112.4940}
  {arXiv:1112.4940 [astro-ph.CO]} \BibitemShut {NoStop}%
\bibitem [{\citenamefont {Palanque-Delabrouille}\ \emph
  {et~al.}(2015)\citenamefont {Palanque-Delabrouille} \emph
  {et~al.}}]{Palanque-Delabrouille:2014jca}%
  \BibitemOpen
  \bibfield  {author} {\bibinfo {author} {\bibfnamefont {N.}~\bibnamefont
  {Palanque-Delabrouille}} \emph {et~al.},\ }\href {\doibase
  10.1088/1475-7516/2015/02/045} {\bibfield  {journal} {\bibinfo  {journal}
  {JCAP}\ }\textbf {\bibinfo {volume} {1502}},\ \bibinfo {pages} {045}
  (\bibinfo {year} {2015})},\ \Eprint {http://arxiv.org/abs/1410.7244}
  {arXiv:1410.7244 [astro-ph.CO]} \BibitemShut {NoStop}%
\bibitem [{\citenamefont {Cuesta}\ \emph {et~al.}(2016)\citenamefont {Cuesta},
  \citenamefont {Niro},\ and\ \citenamefont {Verde}}]{Cuesta:2015iho}%
  \BibitemOpen
  \bibfield  {author} {\bibinfo {author} {\bibfnamefont {A.~J.}\ \bibnamefont
  {Cuesta}}, \bibinfo {author} {\bibfnamefont {V.}~\bibnamefont {Niro}}, \ and\
  \bibinfo {author} {\bibfnamefont {L.}~\bibnamefont {Verde}},\ }\href
  {\doibase 10.1016/j.dark.2016.04.005} {\bibfield  {journal} {\bibinfo
  {journal} {Phys. Dark Univ.}\ }\textbf {\bibinfo {volume} {13}},\ \bibinfo
  {pages} {77} (\bibinfo {year} {2016})},\ \Eprint
  {http://arxiv.org/abs/1511.05983} {arXiv:1511.05983 [astro-ph.CO]}
  \BibitemShut {NoStop}%
\bibitem [{\citenamefont {Olive}\ \emph {et~al.}(2014)\citenamefont {Olive}
  \emph {et~al.}}]{Agashe:2014kda}%
  \BibitemOpen
  \bibfield  {author} {\bibinfo {author} {\bibfnamefont {K.~A.}\ \bibnamefont
  {Olive}} \emph {et~al.} (\bibinfo {collaboration} {Particle Data Group}),\
  }\href {\doibase 10.1088/1674-1137/38/9/090001} {\bibfield  {journal}
  {\bibinfo  {journal} {Chin. Phys.}\ }\textbf {\bibinfo {volume} {C38}},\
  \bibinfo {pages} {090001} (\bibinfo {year} {2014})}\BibitemShut {NoStop}%
\bibitem [{\citenamefont {King}\ \emph {et~al.}(2014)\citenamefont {King},
  \citenamefont {Merle}, \citenamefont {Morisi}, \citenamefont {Shimizu},\ and\
  \citenamefont {Tanimoto}}]{King:2014nza}%
  \BibitemOpen
  \bibfield  {author} {\bibinfo {author} {\bibfnamefont {S.~F.}\ \bibnamefont
  {King}}, \bibinfo {author} {\bibfnamefont {A.}~\bibnamefont {Merle}},
  \bibinfo {author} {\bibfnamefont {S.}~\bibnamefont {Morisi}}, \bibinfo
  {author} {\bibfnamefont {Y.}~\bibnamefont {Shimizu}}, \ and\ \bibinfo
  {author} {\bibfnamefont {M.}~\bibnamefont {Tanimoto}},\ }\href {\doibase
  10.1088/1367-2630/16/4/045018} {\bibfield  {journal} {\bibinfo  {journal}
  {New J. Phys.}\ }\textbf {\bibinfo {volume} {16}},\ \bibinfo {pages} {045018}
  (\bibinfo {year} {2014})},\ \Eprint {http://arxiv.org/abs/1402.4271}
  {arXiv:1402.4271 [hep-ph]} \BibitemShut {NoStop}%
\bibitem [{\citenamefont {Sivia}\ and\ \citenamefont
  {Skilling}(2006)}]{sivia2006data}%
  \BibitemOpen
  \bibfield  {author} {\bibinfo {author} {\bibfnamefont {D.}~\bibnamefont
  {Sivia}}\ and\ \bibinfo {author} {\bibfnamefont {J.}~\bibnamefont
  {Skilling}},\ }\href {https://books.google.com/books?id=6O8ZAQAAIAAJ} {\emph
  {\bibinfo {title} {Data analysis: a Bayesian tutorial}}},\ Oxford science
  publications\ (\bibinfo  {publisher} {Oxford University Press},\ \bibinfo
  {year} {2006})\BibitemShut {NoStop}%
\bibitem [{\citenamefont {Simpson}\ \emph {et~al.}(2017)\citenamefont
  {Simpson}, \citenamefont {Jimenez}, \citenamefont {Pena-Garay},\ and\
  \citenamefont {Verde}}]{Simpson:2017qvj}%
  \BibitemOpen
  \bibfield  {author} {\bibinfo {author} {\bibfnamefont {F.}~\bibnamefont
  {Simpson}}, \bibinfo {author} {\bibfnamefont {R.}~\bibnamefont {Jimenez}},
  \bibinfo {author} {\bibfnamefont {C.}~\bibnamefont {Pena-Garay}}, \ and\
  \bibinfo {author} {\bibfnamefont {L.}~\bibnamefont {Verde}},\ }\href
  {\doibase 10.1088/1475-7516/2017/06/029} {\bibfield  {journal} {\bibinfo
  {journal} {JCAP}\ }\textbf {\bibinfo {volume} {1706}},\ \bibinfo {pages}
  {029} (\bibinfo {year} {2017})},\ \Eprint {http://arxiv.org/abs/1703.03425}
  {arXiv:1703.03425 [astro-ph.CO]} \BibitemShut {NoStop}%
\bibitem [{\citenamefont {Schwetz}\ \emph {et~al.}(2017)\citenamefont
  {Schwetz}, \citenamefont {Freese}, \citenamefont {Gerbino}, \citenamefont
  {Giusarma}, \citenamefont {Hannestad}, \citenamefont {Lattanzi},
  \citenamefont {Mena},\ and\ \citenamefont {Vagnozzi}}]{Schwetz:2017fey}%
  \BibitemOpen
  \bibfield  {author} {\bibinfo {author} {\bibfnamefont {T.}~\bibnamefont
  {Schwetz}}, \bibinfo {author} {\bibfnamefont {K.}~\bibnamefont {Freese}},
  \bibinfo {author} {\bibfnamefont {M.}~\bibnamefont {Gerbino}}, \bibinfo
  {author} {\bibfnamefont {E.}~\bibnamefont {Giusarma}}, \bibinfo {author}
  {\bibfnamefont {S.}~\bibnamefont {Hannestad}}, \bibinfo {author}
  {\bibfnamefont {M.}~\bibnamefont {Lattanzi}}, \bibinfo {author}
  {\bibfnamefont {O.}~\bibnamefont {Mena}}, \ and\ \bibinfo {author}
  {\bibfnamefont {S.}~\bibnamefont {Vagnozzi}},\ }\href@noop {} {\  (\bibinfo
  {year} {2017})},\ \Eprint {http://arxiv.org/abs/1703.04585} {arXiv:1703.04585
  [astro-ph.CO]} \BibitemShut {NoStop}%
\bibitem [{\citenamefont {Hall}\ \emph {et~al.}(2000)\citenamefont {Hall},
  \citenamefont {Murayama},\ and\ \citenamefont {Weiner}}]{Hall:1999sn}%
  \BibitemOpen
  \bibfield  {author} {\bibinfo {author} {\bibfnamefont {L.~J.}\ \bibnamefont
  {Hall}}, \bibinfo {author} {\bibfnamefont {H.}~\bibnamefont {Murayama}}, \
  and\ \bibinfo {author} {\bibfnamefont {N.}~\bibnamefont {Weiner}},\ }\href
  {\doibase 10.1103/PhysRevLett.84.2572} {\bibfield  {journal} {\bibinfo
  {journal} {Phys. Rev. Lett.}\ }\textbf {\bibinfo {volume} {84}},\ \bibinfo
  {pages} {2572} (\bibinfo {year} {2000})},\ \Eprint
  {http://arxiv.org/abs/hep-ph/9911341} {arXiv:hep-ph/9911341 [hep-ph]}
  \BibitemShut {NoStop}%
\bibitem [{\citenamefont {Haba}\ and\ \citenamefont
  {Murayama}(2001)}]{Haba:2000be}%
  \BibitemOpen
  \bibfield  {author} {\bibinfo {author} {\bibfnamefont {N.}~\bibnamefont
  {Haba}}\ and\ \bibinfo {author} {\bibfnamefont {H.}~\bibnamefont
  {Murayama}},\ }\href {\doibase 10.1103/PhysRevD.63.053010} {\bibfield
  {journal} {\bibinfo  {journal} {Phys. Rev.}\ }\textbf {\bibinfo {volume}
  {D63}},\ \bibinfo {pages} {053010} (\bibinfo {year} {2001})},\ \Eprint
  {http://arxiv.org/abs/hep-ph/0009174} {arXiv:hep-ph/0009174 [hep-ph]}
  \BibitemShut {NoStop}%
\bibitem [{\citenamefont {de~Gouvea}\ and\ \citenamefont
  {Murayama}(2003)}]{deGouvea:2003xe}%
  \BibitemOpen
  \bibfield  {author} {\bibinfo {author} {\bibfnamefont {A.}~\bibnamefont
  {de~Gouvea}}\ and\ \bibinfo {author} {\bibfnamefont {H.}~\bibnamefont
  {Murayama}},\ }\href {\doibase 10.1016/j.physletb.2003.08.045} {\bibfield
  {journal} {\bibinfo  {journal} {Phys. Lett.}\ }\textbf {\bibinfo {volume}
  {B573}},\ \bibinfo {pages} {94} (\bibinfo {year} {2003})},\ \Eprint
  {http://arxiv.org/abs/hep-ph/0301050} {arXiv:hep-ph/0301050 [hep-ph]}
  \BibitemShut {NoStop}%
\bibitem [{\citenamefont {Hall}\ \emph {et~al.}(2008)\citenamefont {Hall},
  \citenamefont {Salem},\ and\ \citenamefont {Watari}}]{Hall:2007zh}%
  \BibitemOpen
  \bibfield  {author} {\bibinfo {author} {\bibfnamefont {L.~J.}\ \bibnamefont
  {Hall}}, \bibinfo {author} {\bibfnamefont {M.~P.}\ \bibnamefont {Salem}}, \
  and\ \bibinfo {author} {\bibfnamefont {T.}~\bibnamefont {Watari}},\ }\href
  {\doibase 10.1103/PhysRevLett.100.141801} {\bibfield  {journal} {\bibinfo
  {journal} {Phys. Rev. Lett.}\ }\textbf {\bibinfo {volume} {100}},\ \bibinfo
  {pages} {141801} (\bibinfo {year} {2008})},\ \Eprint
  {http://arxiv.org/abs/0707.3444} {arXiv:0707.3444 [hep-ph]} \BibitemShut
  {NoStop}%
\bibitem [{\citenamefont {Hall}\ \emph {et~al.}(2007)\citenamefont {Hall},
  \citenamefont {Salem},\ and\ \citenamefont {Watari}}]{Hall:2007zj}%
  \BibitemOpen
  \bibfield  {author} {\bibinfo {author} {\bibfnamefont {L.~J.}\ \bibnamefont
  {Hall}}, \bibinfo {author} {\bibfnamefont {M.~P.}\ \bibnamefont {Salem}}, \
  and\ \bibinfo {author} {\bibfnamefont {T.}~\bibnamefont {Watari}},\ }\href
  {\doibase 10.1103/PhysRevD.76.093001} {\bibfield  {journal} {\bibinfo
  {journal} {Phys. Rev.}\ }\textbf {\bibinfo {volume} {D76}},\ \bibinfo {pages}
  {093001} (\bibinfo {year} {2007})},\ \Eprint {http://arxiv.org/abs/0707.3446}
  {arXiv:0707.3446 [hep-ph]} \BibitemShut {NoStop}%
\bibitem [{\citenamefont {Hall}\ \emph {et~al.}(2009)\citenamefont {Hall},
  \citenamefont {Salem},\ and\ \citenamefont {Watari}}]{Hall:2008km}%
  \BibitemOpen
  \bibfield  {author} {\bibinfo {author} {\bibfnamefont {L.~J.}\ \bibnamefont
  {Hall}}, \bibinfo {author} {\bibfnamefont {M.~P.}\ \bibnamefont {Salem}}, \
  and\ \bibinfo {author} {\bibfnamefont {T.}~\bibnamefont {Watari}},\ }\href
  {\doibase 10.1103/PhysRevD.79.025010} {\bibfield  {journal} {\bibinfo
  {journal} {Phys. Rev.}\ }\textbf {\bibinfo {volume} {D79}},\ \bibinfo {pages}
  {025010} (\bibinfo {year} {2009})},\ \Eprint {http://arxiv.org/abs/0810.2561}
  {arXiv:0810.2561 [hep-th]} \BibitemShut {NoStop}%
\bibitem [{\citenamefont {de~Gouvea}\ and\ \citenamefont
  {Murayama}(2015)}]{deGouvea:2012ac}%
  \BibitemOpen
  \bibfield  {author} {\bibinfo {author} {\bibfnamefont {A.}~\bibnamefont
  {de~Gouvea}}\ and\ \bibinfo {author} {\bibfnamefont {H.}~\bibnamefont
  {Murayama}},\ }\href {\doibase 10.1016/j.physletb.2015.06.028} {\bibfield
  {journal} {\bibinfo  {journal} {Phys. Lett.}\ }\textbf {\bibinfo {volume}
  {B747}},\ \bibinfo {pages} {479} (\bibinfo {year} {2015})},\ \Eprint
  {http://arxiv.org/abs/1204.1249} {arXiv:1204.1249 [hep-ph]} \BibitemShut
  {NoStop}%
\bibitem [{\citenamefont {Lu}\ and\ \citenamefont
  {Murayama}(2014)}]{Lu:2014cla}%
  \BibitemOpen
  \bibfield  {author} {\bibinfo {author} {\bibfnamefont {X.}~\bibnamefont
  {Lu}}\ and\ \bibinfo {author} {\bibfnamefont {H.}~\bibnamefont {Murayama}},\
  }\href {\doibase 10.1007/JHEP08(2014)101} {\bibfield  {journal} {\bibinfo
  {journal} {JHEP}\ }\textbf {\bibinfo {volume} {08}},\ \bibinfo {pages} {101}
  (\bibinfo {year} {2014})},\ \Eprint {http://arxiv.org/abs/1405.0547}
  {arXiv:1405.0547 [hep-ph]} \BibitemShut {NoStop}%
\bibitem [{\citenamefont {Fortin}\ \emph {et~al.}(2016)\citenamefont {Fortin},
  \citenamefont {Giasson},\ and\ \citenamefont {Marleau}}]{Fortin:2016zyf}%
  \BibitemOpen
  \bibfield  {author} {\bibinfo {author} {\bibfnamefont {J.-F.}\ \bibnamefont
  {Fortin}}, \bibinfo {author} {\bibfnamefont {N.}~\bibnamefont {Giasson}}, \
  and\ \bibinfo {author} {\bibfnamefont {L.}~\bibnamefont {Marleau}},\ }\href
  {\doibase 10.1103/PhysRevD.94.115004} {\bibfield  {journal} {\bibinfo
  {journal} {Phys. Rev.}\ }\textbf {\bibinfo {volume} {D94}},\ \bibinfo {pages}
  {115004} (\bibinfo {year} {2016})},\ \Eprint
  {http://arxiv.org/abs/1609.08581} {arXiv:1609.08581 [hep-ph]} \BibitemShut
  {NoStop}%
\bibitem [{\citenamefont {Babu}\ \emph {et~al.}(2017)\citenamefont {Babu},
  \citenamefont {Khanov},\ and\ \citenamefont {Saad}}]{Babu:2016aro}%
  \BibitemOpen
  \bibfield  {author} {\bibinfo {author} {\bibfnamefont {K.~S.}\ \bibnamefont
  {Babu}}, \bibinfo {author} {\bibfnamefont {A.}~\bibnamefont {Khanov}}, \ and\
  \bibinfo {author} {\bibfnamefont {S.}~\bibnamefont {Saad}},\ }\href {\doibase
  10.1103/PhysRevD.95.055014} {\bibfield  {journal} {\bibinfo  {journal} {Phys.
  Rev.}\ }\textbf {\bibinfo {volume} {D95}},\ \bibinfo {pages} {055014}
  (\bibinfo {year} {2017})},\ \Eprint {http://arxiv.org/abs/1612.07787}
  {arXiv:1612.07787 [hep-ph]} \BibitemShut {NoStop}%
\bibitem [{\citenamefont {Fortin}\ \emph {et~al.}(2017)\citenamefont {Fortin},
  \citenamefont {Giasson},\ and\ \citenamefont {Marleau}}]{Fortin:2017iiw}%
  \BibitemOpen
  \bibfield  {author} {\bibinfo {author} {\bibfnamefont {J.-F.}\ \bibnamefont
  {Fortin}}, \bibinfo {author} {\bibfnamefont {N.}~\bibnamefont {Giasson}}, \
  and\ \bibinfo {author} {\bibfnamefont {L.}~\bibnamefont {Marleau}},\ }\href
  {\doibase 10.1007/JHEP04(2017)131} {\bibfield  {journal} {\bibinfo  {journal}
  {JHEP}\ }\textbf {\bibinfo {volume} {04}},\ \bibinfo {pages} {131} (\bibinfo
  {year} {2017})},\ \Eprint {http://arxiv.org/abs/1702.07273} {arXiv:1702.07273
  [hep-ph]} \BibitemShut {NoStop}%
\bibitem [{\citenamefont {Susskind}(2003)}]{Susskind:2003kw}%
  \BibitemOpen
  \bibfield  {author} {\bibinfo {author} {\bibfnamefont {L.}~\bibnamefont
  {Susskind}},\ }\href@noop {} {\  (\bibinfo {year} {2003})},\ \Eprint
  {http://arxiv.org/abs/hep-th/0302219} {arXiv:hep-th/0302219 [hep-th]}
  \BibitemShut {NoStop}%
\bibitem [{\citenamefont {Bergstrom}\ \emph {et~al.}(2014)\citenamefont
  {Bergstrom}, \citenamefont {Meloni},\ and\ \citenamefont
  {Merlo}}]{Bergstrom:2014owa}%
  \BibitemOpen
  \bibfield  {author} {\bibinfo {author} {\bibfnamefont {J.}~\bibnamefont
  {Bergstrom}}, \bibinfo {author} {\bibfnamefont {D.}~\bibnamefont {Meloni}}, \
  and\ \bibinfo {author} {\bibfnamefont {L.}~\bibnamefont {Merlo}},\ }\href
  {\doibase 10.1103/PhysRevD.89.093021} {\bibfield  {journal} {\bibinfo
  {journal} {Phys. Rev.}\ }\textbf {\bibinfo {volume} {D89}},\ \bibinfo {pages}
  {093021} (\bibinfo {year} {2014})},\ \Eprint {http://arxiv.org/abs/1403.4528}
  {arXiv:1403.4528 [hep-ph]} \BibitemShut {NoStop}%
\bibitem [{\citenamefont {Hannestad}\ and\ \citenamefont
  {Tram}(2017)}]{Hannestad:2017ypp}%
  \BibitemOpen
  \bibfield  {author} {\bibinfo {author} {\bibfnamefont {S.}~\bibnamefont
  {Hannestad}}\ and\ \bibinfo {author} {\bibfnamefont {T.}~\bibnamefont
  {Tram}},\ }\href@noop {} {\  (\bibinfo {year} {2017})},\ \Eprint
  {http://arxiv.org/abs/1710.08899} {arXiv:1710.08899 [astro-ph.CO]}
  \BibitemShut {NoStop}%
\bibitem [{\citenamefont {Di~Valentino}\ \emph {et~al.}(2016)\citenamefont
  {Di~Valentino}, \citenamefont {Giusarma}, \citenamefont {Mena}, \citenamefont
  {Melchiorri},\ and\ \citenamefont {Silk}}]{DiValentino:2015sam}%
  \BibitemOpen
  \bibfield  {author} {\bibinfo {author} {\bibfnamefont {E.}~\bibnamefont
  {Di~Valentino}}, \bibinfo {author} {\bibfnamefont {E.}~\bibnamefont
  {Giusarma}}, \bibinfo {author} {\bibfnamefont {O.}~\bibnamefont {Mena}},
  \bibinfo {author} {\bibfnamefont {A.}~\bibnamefont {Melchiorri}}, \ and\
  \bibinfo {author} {\bibfnamefont {J.}~\bibnamefont {Silk}},\ }\href {\doibase
  10.1103/PhysRevD.93.083527} {\bibfield  {journal} {\bibinfo  {journal} {Phys.
  Rev.}\ }\textbf {\bibinfo {volume} {D93}},\ \bibinfo {pages} {083527}
  (\bibinfo {year} {2016})},\ \Eprint {http://arxiv.org/abs/1511.00975}
  {arXiv:1511.00975 [astro-ph.CO]} \BibitemShut {NoStop}%
\bibitem [{\citenamefont {Hannestad}\ and\ \citenamefont
  {Schwetz}(2016)}]{Hannestad:2016fog}%
  \BibitemOpen
  \bibfield  {author} {\bibinfo {author} {\bibfnamefont {S.}~\bibnamefont
  {Hannestad}}\ and\ \bibinfo {author} {\bibfnamefont {T.}~\bibnamefont
  {Schwetz}},\ }\href {\doibase 10.1088/1475-7516/2016/11/035} {\bibfield
  {journal} {\bibinfo  {journal} {JCAP}\ }\textbf {\bibinfo {volume} {1611}},\
  \bibinfo {pages} {035} (\bibinfo {year} {2016})},\ \Eprint
  {http://arxiv.org/abs/1606.04691} {arXiv:1606.04691 [astro-ph.CO]}
  \BibitemShut {NoStop}%
\bibitem [{\citenamefont {Gerbino}\ \emph {et~al.}(2016)\citenamefont
  {Gerbino}, \citenamefont {Lattanzi}, \citenamefont {Mena},\ and\
  \citenamefont {Freese}}]{Gerbino:2016ehw}%
  \BibitemOpen
  \bibfield  {author} {\bibinfo {author} {\bibfnamefont {M.}~\bibnamefont
  {Gerbino}}, \bibinfo {author} {\bibfnamefont {M.}~\bibnamefont {Lattanzi}},
  \bibinfo {author} {\bibfnamefont {O.}~\bibnamefont {Mena}}, \ and\ \bibinfo
  {author} {\bibfnamefont {K.}~\bibnamefont {Freese}},\ }\href@noop {} {\
  (\bibinfo {year} {2016})},\ \Eprint {http://arxiv.org/abs/1611.07847}
  {arXiv:1611.07847 [astro-ph.CO]} \BibitemShut {NoStop}%
\bibitem [{\citenamefont {Giusarma}\ \emph {et~al.}(2016)\citenamefont
  {Giusarma}, \citenamefont {Gerbino}, \citenamefont {Mena}, \citenamefont
  {Vagnozzi}, \citenamefont {Ho},\ and\ \citenamefont
  {Freese}}]{Giusarma:2016phn}%
  \BibitemOpen
  \bibfield  {author} {\bibinfo {author} {\bibfnamefont {E.}~\bibnamefont
  {Giusarma}}, \bibinfo {author} {\bibfnamefont {M.}~\bibnamefont {Gerbino}},
  \bibinfo {author} {\bibfnamefont {O.}~\bibnamefont {Mena}}, \bibinfo {author}
  {\bibfnamefont {S.}~\bibnamefont {Vagnozzi}}, \bibinfo {author}
  {\bibfnamefont {S.}~\bibnamefont {Ho}}, \ and\ \bibinfo {author}
  {\bibfnamefont {K.}~\bibnamefont {Freese}},\ }\href {\doibase
  10.1103/PhysRevD.94.083522} {\bibfield  {journal} {\bibinfo  {journal} {Phys.
  Rev.}\ }\textbf {\bibinfo {volume} {D94}},\ \bibinfo {pages} {083522}
  (\bibinfo {year} {2016})},\ \Eprint {http://arxiv.org/abs/1605.04320}
  {arXiv:1605.04320 [astro-ph.CO]} \BibitemShut {NoStop}%
\bibitem [{\citenamefont {Gerbino}\ \emph {et~al.}(2017)\citenamefont
  {Gerbino}, \citenamefont {Freese}, \citenamefont {Vagnozzi}, \citenamefont
  {Lattanzi}, \citenamefont {Mena}, \citenamefont {Giusarma},\ and\
  \citenamefont {Ho}}]{Gerbino:2016sgw}%
  \BibitemOpen
  \bibfield  {author} {\bibinfo {author} {\bibfnamefont {M.}~\bibnamefont
  {Gerbino}}, \bibinfo {author} {\bibfnamefont {K.}~\bibnamefont {Freese}},
  \bibinfo {author} {\bibfnamefont {S.}~\bibnamefont {Vagnozzi}}, \bibinfo
  {author} {\bibfnamefont {M.}~\bibnamefont {Lattanzi}}, \bibinfo {author}
  {\bibfnamefont {O.}~\bibnamefont {Mena}}, \bibinfo {author} {\bibfnamefont
  {E.}~\bibnamefont {Giusarma}}, \ and\ \bibinfo {author} {\bibfnamefont
  {S.}~\bibnamefont {Ho}},\ }\href {\doibase 10.1103/PhysRevD.95.043512}
  {\bibfield  {journal} {\bibinfo  {journal} {Phys. Rev.}\ }\textbf {\bibinfo
  {volume} {D95}},\ \bibinfo {pages} {043512} (\bibinfo {year} {2017})},\
  \Eprint {http://arxiv.org/abs/1610.08830} {arXiv:1610.08830 [astro-ph.CO]}
  \BibitemShut {NoStop}%
\bibitem [{\citenamefont {Couchot}\ \emph {et~al.}(2017)\citenamefont
  {Couchot}, \citenamefont {Henrot-Versill{\'e}}, \citenamefont {Perdereau},
  \citenamefont {Plaszczynski}, \citenamefont {Rouill{\'e} D.~'orfeuil},
  \citenamefont {Spinelli},\ and\ \citenamefont {Tristram}}]{Couchot:2017pvz}%
  \BibitemOpen
  \bibfield  {author} {\bibinfo {author} {\bibfnamefont {F.}~\bibnamefont
  {Couchot}}, \bibinfo {author} {\bibfnamefont {S.}~\bibnamefont
  {Henrot-Versill{\'e}}}, \bibinfo {author} {\bibfnamefont {O.}~\bibnamefont
  {Perdereau}}, \bibinfo {author} {\bibfnamefont {S.}~\bibnamefont
  {Plaszczynski}}, \bibinfo {author} {\bibfnamefont {B.}~\bibnamefont
  {Rouill{\'e} D.~'orfeuil}}, \bibinfo {author} {\bibfnamefont
  {M.}~\bibnamefont {Spinelli}}, \ and\ \bibinfo {author} {\bibfnamefont
  {M.}~\bibnamefont {Tristram}},\ }\href@noop {} {\  (\bibinfo {year}
  {2017})},\ \Eprint {http://arxiv.org/abs/1703.10829} {arXiv:1703.10829
  [astro-ph.CO]} \BibitemShut {NoStop}%
\bibitem [{\citenamefont {Vagnozzi}\ \emph {et~al.}(2017)\citenamefont
  {Vagnozzi}, \citenamefont {Giusarma}, \citenamefont {Mena}, \citenamefont
  {Freese}, \citenamefont {Gerbino}, \citenamefont {Ho},\ and\ \citenamefont
  {Lattanzi}}]{Vagnozzi:2017ovm}%
  \BibitemOpen
  \bibfield  {author} {\bibinfo {author} {\bibfnamefont {S.}~\bibnamefont
  {Vagnozzi}}, \bibinfo {author} {\bibfnamefont {E.}~\bibnamefont {Giusarma}},
  \bibinfo {author} {\bibfnamefont {O.}~\bibnamefont {Mena}}, \bibinfo {author}
  {\bibfnamefont {K.}~\bibnamefont {Freese}}, \bibinfo {author} {\bibfnamefont
  {M.}~\bibnamefont {Gerbino}}, \bibinfo {author} {\bibfnamefont
  {S.}~\bibnamefont {Ho}}, \ and\ \bibinfo {author} {\bibfnamefont
  {M.}~\bibnamefont {Lattanzi}},\ }\href@noop {} {\  (\bibinfo {year}
  {2017})},\ \Eprint {http://arxiv.org/abs/1701.08172} {arXiv:1701.08172
  [astro-ph.CO]} \BibitemShut {NoStop}%
\bibitem [{\citenamefont {Jeffreys}(1946)}]{Jeffreys:1946}%
  \BibitemOpen
  \bibfield  {author} {\bibinfo {author} {\bibfnamefont {H.}~\bibnamefont
  {Jeffreys}},\ }\href@noop {} {\bibfield  {journal} {\bibinfo  {journal}
  {Proceedings of the Royal Society of London. Series A, Mathematical and
  Physical Sciences}\ }\textbf {\bibinfo {volume} {186}},\ \bibinfo {pages}
  {453} (\bibinfo {year} {1946})}\BibitemShut {NoStop}%
\bibitem [{\citenamefont {Kass}\ and\ \citenamefont
  {Wasserman}(1996)}]{Kass:1996}%
  \BibitemOpen
  \bibfield  {author} {\bibinfo {author} {\bibfnamefont {R.~E.}\ \bibnamefont
  {Kass}}\ and\ \bibinfo {author} {\bibfnamefont {L.}~\bibnamefont
  {Wasserman}},\ }\href@noop {} {\bibfield  {journal} {\bibinfo  {journal}
  {Journal of the American Statistical Association}\ }\textbf {\bibinfo
  {volume} {91}},\ \bibinfo {pages} {1343} (\bibinfo {year}
  {1996})}\BibitemShut {NoStop}%
\bibitem [{\citenamefont {{Greg W. Anderson, Alice Guionnet, and Ofer
  Zeitouni}}(2009)}]{RandMat:2009}%
  \BibitemOpen
  \bibfield  {author} {\bibinfo {author} {\bibnamefont {{Greg W. Anderson,
  Alice Guionnet, and Ofer Zeitouni}}},\ }\href@noop {} {\emph {\bibinfo
  {title} {{An Introduction to Random Matrices}}}},\ Cambridge Studies in
  Advanced Mathematics\ (\bibinfo  {publisher} {Cambridge University Press},\
  \bibinfo {year} {2009})\BibitemShut {NoStop}%
\bibitem [{\citenamefont {Minkowski}(1977)}]{Minkowski:1977sc}%
  \BibitemOpen
  \bibfield  {author} {\bibinfo {author} {\bibfnamefont {P.}~\bibnamefont
  {Minkowski}},\ }\href {\doibase 10.1016/0370-2693(77)90435-X} {\bibfield
  {journal} {\bibinfo  {journal} {Phys.Lett.}\ }\textbf {\bibinfo {volume}
  {B67}},\ \bibinfo {pages} {421} (\bibinfo {year} {1977})}\BibitemShut
  {NoStop}%
\bibitem [{\citenamefont {Mohapatra}\ and\ \citenamefont
  {Senjanovic}(1980)}]{Mohapatra:1979ia}%
  \BibitemOpen
  \bibfield  {author} {\bibinfo {author} {\bibfnamefont {R.~N.}\ \bibnamefont
  {Mohapatra}}\ and\ \bibinfo {author} {\bibfnamefont {G.}~\bibnamefont
  {Senjanovic}},\ }\href {\doibase 10.1103/PhysRevLett.44.912} {\bibfield
  {journal} {\bibinfo  {journal} {Phys.Rev.Lett.}\ }\textbf {\bibinfo {volume}
  {44}},\ \bibinfo {pages} {912} (\bibinfo {year} {1980})}\BibitemShut
  {NoStop}%
\bibitem [{\citenamefont {Gell-Mann}\ \emph {et~al.}(1979)\citenamefont
  {Gell-Mann}, \citenamefont {Ramond},\ and\ \citenamefont
  {Slansky}}]{GellMann:1980vs}%
  \BibitemOpen
  \bibfield  {author} {\bibinfo {author} {\bibfnamefont {M.}~\bibnamefont
  {Gell-Mann}}, \bibinfo {author} {\bibfnamefont {P.}~\bibnamefont {Ramond}}, \
  and\ \bibinfo {author} {\bibfnamefont {R.}~\bibnamefont {Slansky}},\
  }\bibfield  {booktitle} {\emph {\bibinfo {booktitle} {{Supergravity Workshop
  Stony Brook, New York, September 27-28, 1979}}},\ }\href@noop {} {\bibfield
  {journal} {\bibinfo  {journal} {Conf. Proc.}\ }\textbf {\bibinfo {volume}
  {C790927}},\ \bibinfo {pages} {315} (\bibinfo {year} {1979})},\ \Eprint
  {http://arxiv.org/abs/1306.4669} {arXiv:1306.4669 [hep-th]} \BibitemShut
  {NoStop}%
\bibitem [{\citenamefont {Yanagida}(1980)}]{Yanagida:1980xy}%
  \BibitemOpen
  \bibfield  {author} {\bibinfo {author} {\bibfnamefont {T.}~\bibnamefont
  {Yanagida}},\ }\href@noop {} {\bibfield  {journal} {\bibinfo  {journal}
  {Prog.Theor.Phys.}\ }\textbf {\bibinfo {volume} {64}},\ \bibinfo {pages}
  {1103} (\bibinfo {year} {1980})}\BibitemShut {NoStop}%
\bibitem [{\citenamefont {Mohapatra}\ and\ \citenamefont
  {Senjanovic}(1981)}]{Mohapatra:1980yp}%
  \BibitemOpen
  \bibfield  {author} {\bibinfo {author} {\bibfnamefont {R.~N.}\ \bibnamefont
  {Mohapatra}}\ and\ \bibinfo {author} {\bibfnamefont {G.}~\bibnamefont
  {Senjanovic}},\ }\href {\doibase 10.1103/PhysRevD.23.165} {\bibfield
  {journal} {\bibinfo  {journal} {Phys.Rev.}\ }\textbf {\bibinfo {volume}
  {D23}},\ \bibinfo {pages} {165} (\bibinfo {year} {1981})}\BibitemShut
  {NoStop}%
\bibitem [{\citenamefont {Schechter}\ and\ \citenamefont
  {Valle}(1980)}]{Schechter:1980gr}%
  \BibitemOpen
  \bibfield  {author} {\bibinfo {author} {\bibfnamefont {J.}~\bibnamefont
  {Schechter}}\ and\ \bibinfo {author} {\bibfnamefont {J.}~\bibnamefont
  {Valle}},\ }\href {\doibase 10.1103/PhysRevD.22.2227} {\bibfield  {journal}
  {\bibinfo  {journal} {Phys.Rev.}\ }\textbf {\bibinfo {volume} {D22}},\
  \bibinfo {pages} {2227} (\bibinfo {year} {1980})}\BibitemShut {NoStop}%
\bibitem [{\citenamefont {Dreiner}\ \emph {et~al.}(2010)\citenamefont
  {Dreiner}, \citenamefont {Haber},\ and\ \citenamefont
  {Martin}}]{Dreiner:2008tw}%
  \BibitemOpen
  \bibfield  {author} {\bibinfo {author} {\bibfnamefont {H.~K.}\ \bibnamefont
  {Dreiner}}, \bibinfo {author} {\bibfnamefont {H.~E.}\ \bibnamefont {Haber}},
  \ and\ \bibinfo {author} {\bibfnamefont {S.~P.}\ \bibnamefont {Martin}},\
  }\href {\doibase 10.1016/j.physrep.2010.05.002} {\bibfield  {journal}
  {\bibinfo  {journal} {Phys.Rept.}\ }\textbf {\bibinfo {volume} {494}},\
  \bibinfo {pages} {1} (\bibinfo {year} {2010})},\ \Eprint
  {http://arxiv.org/abs/0812.1594} {arXiv:0812.1594 [hep-ph]} \BibitemShut
  {NoStop}%
\bibitem [{\citenamefont {Capozzi}\ \emph {et~al.}(2016)\citenamefont
  {Capozzi}, \citenamefont {Lisi}, \citenamefont {Marrone}, \citenamefont
  {Montanino},\ and\ \citenamefont {Palazzo}}]{Capozzi:2016rtj}%
  \BibitemOpen
  \bibfield  {author} {\bibinfo {author} {\bibfnamefont {F.}~\bibnamefont
  {Capozzi}}, \bibinfo {author} {\bibfnamefont {E.}~\bibnamefont {Lisi}},
  \bibinfo {author} {\bibfnamefont {A.}~\bibnamefont {Marrone}}, \bibinfo
  {author} {\bibfnamefont {D.}~\bibnamefont {Montanino}}, \ and\ \bibinfo
  {author} {\bibfnamefont {A.}~\bibnamefont {Palazzo}},\ }\href {\doibase
  10.1016/j.nuclphysb.2016.02.016} {\bibfield  {journal} {\bibinfo  {journal}
  {Nucl. Phys.}\ }\textbf {\bibinfo {volume} {B908}},\ \bibinfo {pages} {218}
  (\bibinfo {year} {2016})},\ \Eprint {http://arxiv.org/abs/1601.07777}
  {arXiv:1601.07777 [hep-ph]} \BibitemShut {NoStop}%
\bibitem [{\citenamefont {Foreman-Mackey}\ \emph {et~al.}(2013)\citenamefont
  {Foreman-Mackey}, \citenamefont {Hogg}, \citenamefont {Lang},\ and\
  \citenamefont {Goodman}}]{ForemanMackey:2012ig}%
  \BibitemOpen
  \bibfield  {author} {\bibinfo {author} {\bibfnamefont {D.}~\bibnamefont
  {Foreman-Mackey}}, \bibinfo {author} {\bibfnamefont {D.~W.}\ \bibnamefont
  {Hogg}}, \bibinfo {author} {\bibfnamefont {D.}~\bibnamefont {Lang}}, \ and\
  \bibinfo {author} {\bibfnamefont {J.}~\bibnamefont {Goodman}},\ }\href
  {\doibase 10.1086/670067} {\bibfield  {journal} {\bibinfo  {journal} {Publ.
  Astron. Soc. Pac.}\ }\textbf {\bibinfo {volume} {125}},\ \bibinfo {pages}
  {306} (\bibinfo {year} {2013})},\ \Eprint {http://arxiv.org/abs/1202.3665}
  {arXiv:1202.3665 [astro-ph.IM]} \BibitemShut {NoStop}%
\bibitem [{get()}]{getDist}%
  \BibitemOpen
  \href@noop {} {\enquote {\bibinfo {title} {Getdist website},}\ }\bibinfo
  {howpublished} {\url{https://github.com/cmbant/getdist}}\BibitemShut
  {NoStop}%
\bibitem [{\citenamefont {Heavens}\ \emph {et~al.}(2017)\citenamefont
  {Heavens}, \citenamefont {Fantaye}, \citenamefont {Mootoovaloo},
  \citenamefont {Eggers}, \citenamefont {Hosenie}, \citenamefont {Kroon},\ and\
  \citenamefont {Sellentin}}]{Heavens:2017afc}%
  \BibitemOpen
  \bibfield  {author} {\bibinfo {author} {\bibfnamefont {A.}~\bibnamefont
  {Heavens}}, \bibinfo {author} {\bibfnamefont {Y.}~\bibnamefont {Fantaye}},
  \bibinfo {author} {\bibfnamefont {A.}~\bibnamefont {Mootoovaloo}}, \bibinfo
  {author} {\bibfnamefont {H.}~\bibnamefont {Eggers}}, \bibinfo {author}
  {\bibfnamefont {Z.}~\bibnamefont {Hosenie}}, \bibinfo {author} {\bibfnamefont
  {S.}~\bibnamefont {Kroon}}, \ and\ \bibinfo {author} {\bibfnamefont
  {E.}~\bibnamefont {Sellentin}},\ }\href@noop {} {\  (\bibinfo {year}
  {2017})},\ \Eprint {http://arxiv.org/abs/1704.03472} {arXiv:1704.03472
  [stat.CO]} \BibitemShut {NoStop}%
\bibitem [{\citenamefont {Feroz}\ \emph {et~al.}(2009)\citenamefont {Feroz},
  \citenamefont {Hobson},\ and\ \citenamefont {Bridges}}]{Feroz:2008xx}%
  \BibitemOpen
  \bibfield  {author} {\bibinfo {author} {\bibfnamefont {F.}~\bibnamefont
  {Feroz}}, \bibinfo {author} {\bibfnamefont {M.~P.}\ \bibnamefont {Hobson}}, \
  and\ \bibinfo {author} {\bibfnamefont {M.}~\bibnamefont {Bridges}},\ }\href
  {\doibase 10.1111/j.1365-2966.2009.14548.x} {\bibfield  {journal} {\bibinfo
  {journal} {Mon. Not. Roy. Astron. Soc.}\ }\textbf {\bibinfo {volume} {398}},\
  \bibinfo {pages} {1601} (\bibinfo {year} {2009})},\ \Eprint
  {http://arxiv.org/abs/0809.3437} {arXiv:0809.3437 [astro-ph]} \BibitemShut
  {NoStop}%
\bibitem [{\citenamefont {Buchner}\ \emph {et~al.}(2014)\citenamefont
  {Buchner}, \citenamefont {Georgakakis}, \citenamefont {Nandra}, \citenamefont
  {Hsu}, \citenamefont {Rangel}, \citenamefont {Brightman}, \citenamefont
  {Merloni}, \citenamefont {Salvato}, \citenamefont {Donley},\ and\
  \citenamefont {Kocevski}}]{Buchner:2014nha}%
  \BibitemOpen
  \bibfield  {author} {\bibinfo {author} {\bibfnamefont {J.}~\bibnamefont
  {Buchner}}, \bibinfo {author} {\bibfnamefont {A.}~\bibnamefont
  {Georgakakis}}, \bibinfo {author} {\bibfnamefont {K.}~\bibnamefont {Nandra}},
  \bibinfo {author} {\bibfnamefont {L.}~\bibnamefont {Hsu}}, \bibinfo {author}
  {\bibfnamefont {C.}~\bibnamefont {Rangel}}, \bibinfo {author} {\bibfnamefont
  {M.}~\bibnamefont {Brightman}}, \bibinfo {author} {\bibfnamefont
  {A.}~\bibnamefont {Merloni}}, \bibinfo {author} {\bibfnamefont
  {M.}~\bibnamefont {Salvato}}, \bibinfo {author} {\bibfnamefont
  {J.}~\bibnamefont {Donley}}, \ and\ \bibinfo {author} {\bibfnamefont
  {D.}~\bibnamefont {Kocevski}},\ }\href {\doibase 10.1051/0004-6361/201322971}
  {\bibfield  {journal} {\bibinfo  {journal} {Astron. Astrophys.}\ }\textbf
  {\bibinfo {volume} {564}},\ \bibinfo {pages} {A125} (\bibinfo {year}
  {2014})},\ \Eprint {http://arxiv.org/abs/1402.0004} {arXiv:1402.0004
  [astro-ph.HE]} \BibitemShut {NoStop}%
\end{thebibliography}%

\end{document}